%% file: main.tex
\documentclass[a4paper,UKenglish,cleveref, autoref, thm-restate]{lipics-v2021}
\hideLIPIcs  %uncomment to remove references to LIPIcs series (logo, DOI, ...), e.g. when preparing a pre-final version to be uploaded to arXiv or another public repository

\nolinenumbers

\usepackage{enumitem}
\usepackage{algorithm}
\usepackage{booktabs}
\usepackage[noend]{algpseudocode}
\usepackage{xspace}

\usepackage[labelformat=simple]{subcaption}

\usepackage[font=small,labelfont=bf]{caption}
\usepackage{enumitem}
\title{ParGeo: A Library for Parallel Computational Geometry}

\author{Yiqiu Wang}{MIT CSAIL}{yiqiuw@mit.edu}{[orcid]}{}
\author{Rahul Yesantharao}{MIT CSAIL}{rahuly@mit.edu}{[orcid]}{}
\author{Shangdi Yu}{MIT CSAIL}{shangdiy@mit.edu}{[orcid]}{}
\author{Laxman Dhulipala}{University of Maryland, College Park}{laxman@umd.edu}{[orcid]}{}
\author{Yan Gu}{University of California, Riverside}{ygu@cs.ucr.edu}{[orcid]}{}
\author{Julian Shun}{MIT CSAIL}{jshun@mit.edu}{[orcid]}{}
\authorrunning{Y. Wang, R. Yesantharao, S. Yu, L. Dhulipala, Y. Gu, J. Shun} %TODO mandatory. First: Use abbreviated first/middle names. Second (only in severe cases): Use first author plus 'et al.'

\Copyright{Yiqiu Wang, Rahul Yesantharao, Shangdi Yu, Laxman Dhulipala, Yan Gu, Julian Shun} %TODO mandatory, please use full first names. LIPIcs license is "CC-BY";  http://creativecommons.org/licenses/by/3.0/


\supplementdetails[, , subcategory={}, swhid={}]{Software}{https://github.com/ParAlg/ParGeo} %linktext, cite, and subcategory are optional
\supplementdetails[, , subcategory={}, swhid={}]{Related Paper}{https://arxiv.org/abs/2112.06188} % (let's link the full version of the ParGeo paper)
\EventEditors{Shiri Chechik, Gonzalo Navarro, Eva Rotenberg, and Grzegorz Herman}
\EventNoEds{4}
\EventLongTitle{30th Annual European Symposium on Algorithms (ESA 2022)}
\EventShortTitle{ESA 2022}
\EventAcronym{ESA}
\EventYear{2022}
\EventDate{September 5--9, 2022}
\EventLocation{Berlin/Potsdam, Germany}
\EventLogo{}
\SeriesVolume{244}
\ArticleNo{81}
\newif\iffull
\fulltrue

\begin{document}

\maketitle

\input{macro}
\input{batch-kdtree/macro}

\input{sections/abstract}
\input{sections/intro}
% \input{sections/prelim} %% Merged into other paragraphs
\input{sections/framework}
\input{sections/hull}

\input{sections/seb}

\input{sections/batch-kdtree}
\input{sections/experiments}
\input{sections/conclusion}

\myparagraph{Acknowledgements}
This research is supported by
DOE Early Career Award \#DE-SC0018947,
NSF CAREER Award \#CCF-1845763, Google Faculty Research Award, Google Research Scholar Award, DARPA
SDH Award \#HR0011-18-3-0007, and Applications Driving Architectures
(ADA) Research Center, a JUMP Center co-sponsored by SRC and DARPA.
%%
%% Bibliography
%%

%% Please use bibtex, 

\bibliography{strings,references,batch-kdtree-refs}

\iffull

\appendix

\input{sections/appendix/hull}

% \input{sections/appendix/seb} % Moved to body
\input{sections/appendix/bdl-tree}
\input{sections/appendix/experiments}

\fi

\end{document}

%% file: macro.tex
\def\omitcomments{0}

\if\omitcomments 1

\newcommand{\yiqiu}[1]{}
\newcommand{\laxman}[1]{}
\newcommand{\yan}[1]{}
\newcommand{\julian}[1]{}
\newcommand{\shangdi}[1]{}

\else

\newcommand{\yiqiu}[1]{\noindent
  {\color{blue}{\textbf{Yiqiu: }#1}}}
\newcommand{\laxman}[1]{{\color{purple}{\bf Laxman:} #1}}
\newcommand{\yan}[1]{{\color{purple}{\bf Yan:} #1}}
\newcommand{\julian}[1]{{\color{red}{\bf Julian:} #1}}
\newcommand{\shangdi}[1]{{\color{orange}{\bf Shangdi:} #1}}

\fi

\newcommand{\defn}[1]{\emph{\textbf{#1}}}

\newcommand{\true}{\texttt{true}}
\newcommand{\false}{\texttt{false}}

\newcommand{\codelineskip}{\vspace{.03in}}

\newcommand{\hide}[1]{}
\newcommand{\remove}[1]{}
\newcommand{\etal}{{et al}.\xspace }

\newcommand{\writeMin}{\textbf{WriteMin}}
\newcommand{\pack}{\textbf{ParallelPack}}

\newcommand{\myparagraph}[1]{\vspace{1pt}\noindent {\bf #1.}}

\newcommand{\pargeo}{\textsc{ParGeo}}

\lstset{basicstyle=\small\ttfamily, tabsize=2, escapeinside={@}{@},
columns=flexible, showstringspaces=false, frame=single, numberblanklines=false}
\lstset{literate={<|}{{$\langle$}}1  {|>}{{$\rangle$}}1}
\lstset{language=C++, morekeywords={CAS,and,or,in,each,commit,empty,new,job,taken,entry,GOTO,bool}}
%\lstset{xleftmargin=5.0ex, numbers=left}
%\lstset{caption={\texttt{tuple} and \collect{} algorithms},captionpos=b,label={alg:collect}}
\makeatletter
\lst@Key{countblanklines}{true}[t]%
{\lstKV@SetIf{#1}\lst@ifcountblanklines}

\def\ContinueLineNumber{\lstset{firstnumber=last}}
\def\StartLineAt#1{\lstset{firstnumber=#1}}

\lst@AddToHook{OnEmptyLine}{%
  \lst@ifnumberblanklines\else%
  \lst@ifcountblanklines\else%
  \advance\c@lstnumber-\@ne\relax%
  \fi%
  \fi}
\makeatother

%% file: batch-kdtree/macro.tex
%\newif\iffull
%\fullfalse
%\fulltrue

\newcommand{\kdtree}{$k$d-tree}
\newcommand{\kdtrees}{\kdtree s}
\newcommand{\knn}{$k$-NN}
\newcommand{\logtree}{BDL-tree}
\newcommand{\ourtree}{BDL-tree}
\newcommand{\buildveb}{BuildvEB$_\text{S}$}
\newcommand{\buildvebrecursive}{BuildvEBRecursive$_\text{S}$}

\newcommand{\buildbhl}{Build$_\text{S}$}
\newcommand{\buildbhlrecursive}{BuildRecursive$_\text{S}$}

\newcommand{\eraseS}{Erase$_\text{S}$}
\newcommand{\eraseSrecursive}{EraseRecursive$_\text{S}$}

\newcommand{\insertlog}{Insert$_\text{L}$}
\newcommand{\eraselog}{Erase$_\text{L}$}
\newcommand{\knnlog}{kNN$_\text{L}$}
\newcommand{\knnserial}{kNN$_\text{S}$}

\newcommand{\tablecaption}[1]{\vspace{5pt}\caption{#1}}

\newcommand{\inputtable}[3]{
\begin{table}[]
\centering
\input{#1}
\tablecaption{#2}
\label{#3}
\end{table}
}

\newcommand{\inputsubtable}[3]{
\begin{subtable}{0.5\textwidth}
\centering
\input{#1}
\vspace{-7pt}
\tablecaption{#2}
\label{#3}
\end{subtable}
}

%% file: sections/abstract.tex
\begin{abstract}

%Existing computational geometry libraries such as CGAL, while providing support for many important geometric tasks, do not provide adequate support for multicore parallelism.
This paper presents ParGeo, a multicore library for computational geometry.
ParGeo contains modules for fundamental tasks including $k$d-tree based spatial search, spatial graph generation, and
algorithms in computational geometry.

We focus on three new algorithmic contributions provided in the library.
First, we present a new parallel convex hull algorithm based on a
reservation technique to enable parallel modifications to the
hull. We also provide the first parallel implementations of the
randomized incremental convex hull algorithm as well as a
divide-and-conquer convex hull algorithm in $\mathbb{R}^3$. 
Second, for the smallest enclosing ball problem, we propose a new
sampling-based algorithm to quickly reduce the size of the
data set. We also provide the first parallel implementation of
Welzl's classic algorithm for smallest enclosing ball. 
Third, we present the BDL-tree, a parallel batch-dynamic $k$d-tree that allows for efficient parallel updates and $k$-NN queries over dynamically changing point sets. BDL-trees consist of a log-structured set of $k$d-trees which can be used to efficiently insert, delete, and query batches of points in parallel.

On 36 cores with two-way hyper-threading, our fastest convex hull algorithm achieves up to 44.7x self-relative parallel speedup and up to
559x speedup against the best existing sequential implementation.
Our smallest enclosing ball algorithm using our sampling-based algorithm
achieves up to 27.1x self-relative parallel speedup and up to 178x speedup
against the best existing sequential implementation. Our implementation of the BDL-tree achieves
self-relative parallel speedup of up to 46.1x. Across all of the algorithms in ParGeo, we achieve self-relative parallel speedup of 8.1--46.61x.

\end{abstract}

%% file: sections/intro.tex
\section{Introduction}
\label{sec:intro}

Computational geometry algorithms have important applications in various domains, including computer graphics, robotics, computer vision, and geographic information systems~\cite{deberg2000compgeom,PS1985}. 
It is important to provide users with libraries of efficient computational geometry algorithms that they can easily use in their own higher-level applications.
Furthermore, due to the growing sizes of data sets that need to be processed today, and the ubiquity of parallel (multicore) machines, it is beneficial to use parallel algorithms to speed up computations. 
In this paper, we present the ParGeo library for parallel computational geometry,
which includes a rich set of parallel algorithms for geometric problems and data structures,
including $k$d-trees, $k$-nearest neighbor search, range search,
well-separated pair decomposition, Euclidean minimum spanning tree,
spatial sorting, and geometric clustering.
ParGeo also contains a collection of geometric graph generators, including
$k$-nearest neighbor graphs and various spatial networks.
Algorithms from ParGeo can either run sequentially, or 
run using parallel schedulers such as OpenMP, Cilk, or ParlayLib.

While there exist numerous libraries for computational geometry,
most of them are not designed for parallel processing.
For example, Libigl~\cite{libigl} is a library that specializes in
the construction of discrete differential geometry
operators and finite-element matrices.
However, only some aspects of Libigl take advantage of
parallelism.
In comparison, the algorithms and implementations of ParGeo
are designed for parallelism, and target a different set of problems.
CGAL (Computational Geometry Algorithms Library)~\cite{cgal}
is a well-known library of computational geometry algorithms
that includes a wide range of algorithms, but most implementations are not parallel.
Batista~\etal~\cite{batista2010} targeted a few important
algorithms, including spatial sorting, box intersection, and
Delaunay triangulation for
shared-memory parallel processing, with code in CGAL.
In comparison, ParGeo targets similar classes of problems as CGAL,
but \emph{all} of our implementations are highly parallel.
%We list several other libraries in Appendix~\ref{sec:more-libraries}. 
PMP~\cite{pmp-library}, Cinolib~\cite{cinolib}, and Tetwild~\cite{tetwild} 
are libraries for
polygonal and polyhedron meshes, tackling different problems from ParGeo.
MatGeom~\cite{matgeom} is a library for sequential geometric computing
with MATLAB.
The Problem Based Benchmark Suite~\cite{pbbs,Anderson2022}
is a multicore benchmark suite that has some overlap in algorithms with ParGeo. 
LEDA~\cite{leda} is a library of
data structures and algorithms for sequential combinatorial and geometric processing.
ArborX~\cite{arborx} is a parallel library for spatial search.

In this paper, in addition to providing an overview of work on ParGeo, 
we describe new parallel algorithms implemented in ParGeo for convex hull,
smallest enclosing ball, and batch-dynamic \kdtree{} that we developed.
% Both structures can be used to approximate objects to speed up collision detection and neighbor queries, and to perform path planning~\cite{chang2010efficient,nearchou1994collision,larsson2009adaptive}.
% In addition, the convex hull can be used for mesh generation~\cite{shewchuk1999lecture} and cluster analysis~\cite{liparulo2015}, and the smallest enclosing ball can be used to construct the sphere-tree~\cite{katayama1997sstree} or the bounding-volume hierarchy~\cite{dinas2015literature}.
For convex hull, we develop new parallel algorithms for both $\mathbb{R}^2$ and $\mathbb{R}^3$, where
our key algorithmic novelty is a
reservation technique to enable parallel modifications to the hull.
For smallest enclosing ball,
we propose a new sampling-based algorithm based on Larsson~\etal's~\cite{larsson2016} approach to quickly reduce the size of the data set. We also provide the first parallel implementation of the classic randomized incremental algorithm~\cite{clarkson89random}.
For \kdtrees{},
we develop the \ourtree{}, a new parallel data structure that supports batch-dynamic operations (construction, insertions, and deletions) as well as exact \knn{} queries.
\ourtree{}s consist of a set of exponentially growing \kdtrees{} and perform batched updates in parallel.

To demonstrate the efficiency of our proposed algorithms and library,
we perform a comprehensive set of experiments on synthetic and
real-world geometric data sets, and compare the performance across our parallel implementations as well as optimized sequential baselines.
On 36 cores with two-way hyper-threading, our best convex
hull implementations achieve up to 44.7x (42.8x on average) self-relative speedup and up to
559x (325x on average) speedup against the best existing sequential implementation
for $\mathbb{R}^2$, and up to 24.9x (11.81x on average) self-relative speedup and up to
124x (61.4x on average) speedup against the best existing sequential implementation
for $\mathbb{R}^3$.
Our sampling-based smallest enclosing ball algorithm
achieves up to 27.1x (20.08x on average) self-relative speedup and up to 178x (109x on average) speedup
against the best existing sequential implementation for $\mathbb{R}^2$ and $\mathbb{R}^3$.
% We experimentally evaluate \ourtree{}s by designing a set of benchmarks to compare its performance against the two baseline approaches described above, which we implemented using similar optimizations. First, we perform scalability tests for each of the four main operations, construction, batch insertion, batch deletion, and \knn{} in order to evaluate the scalability of our data structure on many cores.
Our \ourtree{} achieves self-relative speedup of up to $35.4\times$ ($30.0\times$ on average) for construction, up to $35.0\times$ ($28.3\times$ on average) for batch insertion, up to $33.1\times$ ($28.5\times$ on average) for batch deletion, and up to $46.1\times$ ($40.0\times$ on average) for full \knn{}.
% The largest dataset we test consists of 321 million 3-dimensional points. Then, we design a set of benchmarks that perform a mixed set of updates and queries in order to better understand the performance of \ourtree{}s in realistic scenarios. We find that, when faced with a mixed set of batch operations, \ourtree{}s consistently outperforms the two baselines and presents the best option for such a mixed dynamic setting.
Finally, across all implementations in ParGeo, we achieve self-relative parallel speedup of
8.1--46.6x (on average 23.2x).
% The ParGeo library is publicly available, and due to double-blind requirements we will provide the link to the code when our paper is published.
% at \url{https://github.com/ParAlg/ParGeo}.

%%% Summarize contributions

% We summarize our contributions below:
% \begin{itemize}[topsep=1pt,itemsep=0pt,parsep=0pt,leftmargin=15pt]
%   \item The ParGeo library for parallel
%   computational geometry, which includes the problems studied in this paper as well as a collection of other geometric algorithms and data structures (open sourced).
%   \item Optimized parallel randomized incremental and quickhull algorithms using the reservation technique, as well as a divide-and-conquer algorithm, for convex hull in $\mathbb{R}^2$ and $\mathbb{R}^3$.
%   \item A parallel sampling-based algorithm for the smallest enclosing ball problem, and the first parallel implementation of the classic randomized incremental algorithm.
%   \item New parallel batch-dynamic algorithms for the \kdtree{}.
%   % \item A comprehensive experimental study of our implementations on a 36-core machine with two-way hyper-threading, showing speedups of up to 559x for convex hull and up to 178x for smallest enclosing ball compared with the fastest existing sequential implementations.
% \end{itemize}

% \input{batch-kdtree/sections/introduction}

%% file: sections/framework.tex
\section{The ParGeo Library}\label{sec:framework}

\input{sections/framework/overview-diagram}

Our main goal in designing ParGeo was to enable reusable and efficient
parallel implementations of geometric algorithms and data structures.
We present an overview of the modules of ParGeo in Figure~\ref{fig:pargeo-oveview}, highlighting how the modules interact with each other.
% \laxman{You could number the modules in the figure so that you can mention that here, e.g. (1) (2) (3) (4)}\yiqiu{Sounds good -- added} \julian{I think you can just number the 4 modules instead of all 16 algorithms so it's less cluttered} \yiqiu{Oh I see. Will change it later}
ParGeo contains efficient multicore implementations of static and batch-dynamic \kdtree{}s (Module (1)).
The code supports \kdtree{} based spatial search, including
$k$-nearest neighbor and range search.
Our code is optimized for fast \kdtree{} construction by performing the split in parallel (either by spatial median or by object median),
and performing the queries in a data-parallel fashion.
% We also provide a new cache-oblivious algorithm for \kdtree{} construction (Section~\ref{section:single-alg-top}).
%ParGeo also contains a parallel-batch dynamic \kdtree{} using the logarithmic method,
which we will introduce in Section~\ref{sec:ourtree}.

ParGeo contains a module for parallel computational geometry algorithms (Module (2)).
Our \kdtree{} can be used to generate a well-separated pair
decomposition~\cite{callahan95} (WSPD), which can in turn be used to compute the hierarchical DBSCAN~\cite{wang2021emst}, 
ParGeo contains parallel implementations for the bichromatic closest pair, closest pair, convex hull, smallest enclosing ball, and Morton sorting.

% Our paper is going to focus on more on two other problems in later sections, namely
% the convex hull (Section~\ref{sec:hull}) and the smallest enclosing ball (Section~\ref{sec:seb}).

In addition, ParGeo contains a collection of geometric graph generators (Module (3))
for point data sets.
Our \kdtree{}'s \knn{} search is used to generate the \knn{} graph,
and the range search is used to generate the $\beta$-skeleton graph~\cite{KIRKPATRICK1985217}.
Our WSPD generated from the \kdtree{}
can also be used to compute the Euclidean minimum spanning tree~\cite{callahan93,wang2021emst}, and spanners~\cite{callahan95}.
ParGeo also generates the Delaunay graph.

ParGeo contains a point data generator module (Module (4)) for which can generate uniformly
distributed data sets, and clustered data sets of varying densities~\cite{gantao2015}.
These data sets are used for benchmarking the other modules.

As shown in Table~\ref{tab:all-bench}, on a machine with 36 cores with
two-way hyper-threading, ParGeo achieves self-relative parallel speedups of
8.1--46.61x (23.15x on average) on a uniformly distributed
data set, across all of the benchmarks.
In the subsequent sections, we present three new algorithmic contributions
provided in the library.

\input{tables/pargeo-speedup}

%% file: sections/framework/overview-diagram.tex
\begin{figure*}
    \centering
    \includegraphics[width=0.9\textwidth]{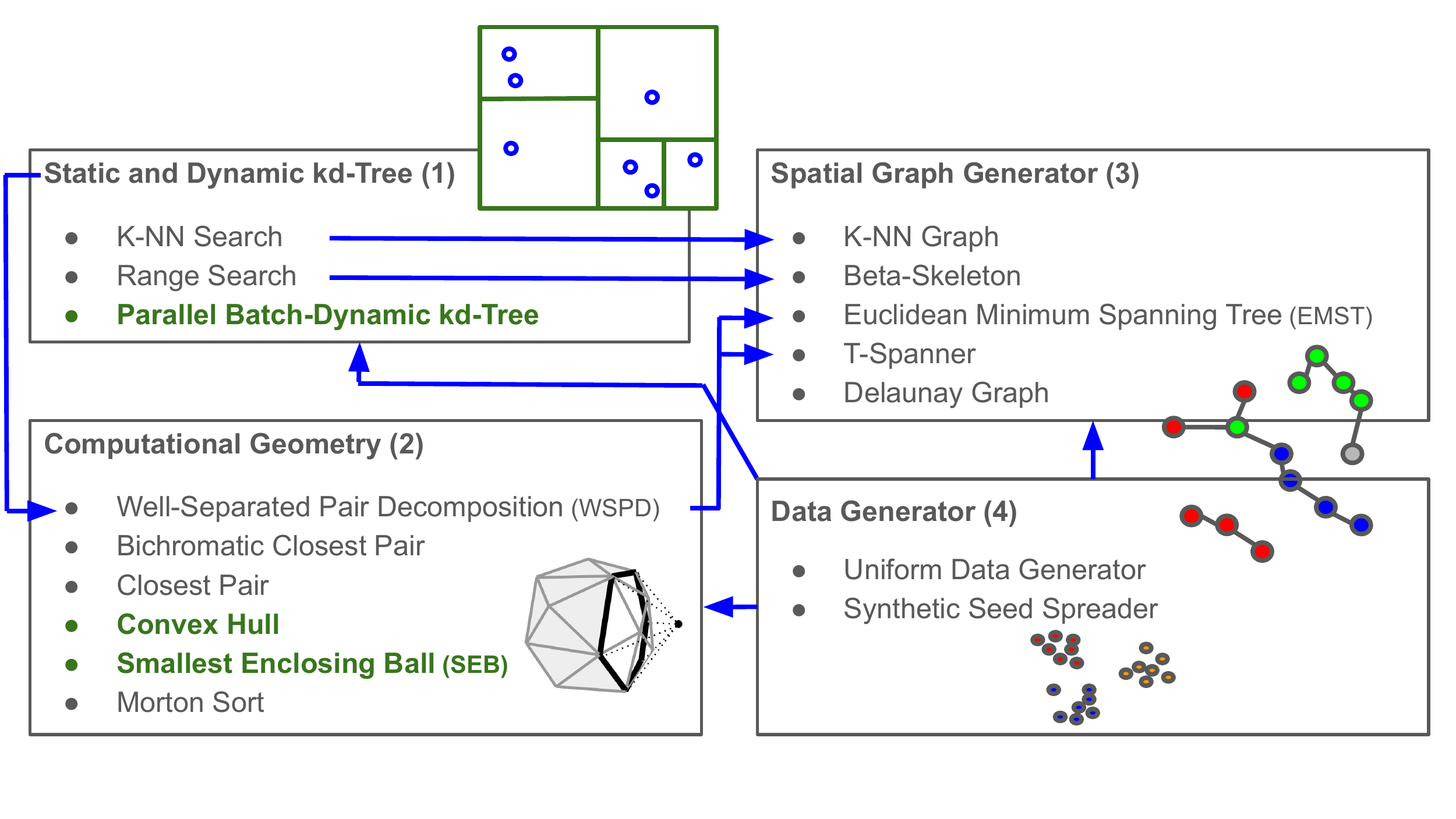}
    \vspace{-10pt}
    \caption{The figure shows an overview of modules in ParGeo. An arrow indicates that a component is used inside another component. In this paper, we present new algorithms and techniques for the modules highlighted in green.} \label{fig:pargeo-oveview}
\end{figure*}

%% file: tables/pargeo-speedup.tex
\begin{table}[t]
\vspace{-3pt}
\footnotesize
\centering
\begin{tabular}{@{}ll|rlrl}
\hline
\multicolumn{2}{l|}{\bf Implementation}              & \multicolumn{1}{c}{$\mathbf{T_1}$} & \multicolumn{1}{l}{$\mathbf{T_{36h}}$} &  \multicolumn{2}{c}{\bf Speedup} \\ \hline
\multicolumn{2}{l|}{\emph{\kdtree{} Build (2d)}}          & 5.51        & 0.43          & 12.70x           \\
\multicolumn{2}{l|}{\emph{\kdtree{} Build (5d)}}          & 8.39        & 0.89          & 9.40x            \\
\multicolumn{2}{l|}{\emph{\kdtree{} $k$-NN (2d)}}            & 31.45       & 0.68          & 46.34x           \\
\multicolumn{2}{l|}{\emph{\kdtree{} Range Search (2d)}}    & 17.14       & 0.37          & 46.61x           \\
% \multicolumn{2}{l|}{\kdtree{}-orthSearch (2d)}     & 16.96       & 0.36          & 46.51x           \\
%\multicolumn{2}{l|}{\kdtree{}-BCP (2d)}            & 0.91        & --             & --        
%\\ 
\hline
\multicolumn{2}{l|}{\emph{Batch-dynamic \kdtree{} Construction (5d)}} & 6.70        & 0.60          & 10.70x            \\
\multicolumn{2}{l|}{\emph{Batch-dynamic \kdtree{} Insert (5d)}} & 8.80        & 1.10          & 8.10x            \\
\multicolumn{2}{l|}{\emph{Batch-dynamic \kdtree{} Delete (5d)}} & 29.20        & 1.20          & 23.90x            \\ \hline
\multicolumn{2}{l|}{\emph{WSPD (2d)}}                  & 6.72        & 0.24          & 27.63x           \\ \hline
\multicolumn{2}{l|}{\emph{EMST (2d)}}                  & 33.02       & 1.58          & 20.86x           \\ \hline
\multicolumn{2}{l|}{\emph{Convex Hull (2d)}}           & 0.38        & 0.0088        & 43.13x           \\
\multicolumn{2}{l|}{\emph{Convex Hull (3d)}}           & 2.36        & 0.097         & 24.36x           \\ \hline
\multicolumn{2}{l|}{\emph{Smallest Enclosing Ball (2d)}}  & 0.053       & 0.0033        & 16.30x           \\
\multicolumn{2}{l|}{\emph{Smallest Enclosing Ball (5d)}}  & 0.13        & 0.014         & 9.54x            \\ \hline
\multicolumn{2}{l|}{\emph{Closest Pair (2d)}}            & 10.35        & 0.52          & 19.90x           \\
\multicolumn{2}{l|}{\emph{Closest Pair (3d)}}            & 28.00        & 2.32          & 12.07x            \\ \hline
\multicolumn{2}{l|}{\emph{$k$-NN Graph (2d)}}             & 37.89       & 1.46          & 25.99x           \\
\multicolumn{2}{l|}{\emph{Delaunay Graph (2d)}}        & 55.91       & 2.03          & 27.53x           \\
\multicolumn{2}{l|}{\emph{Gabriel Graph (2d)}}         & 59.61       & 1.99          & 29.99x           \\
\multicolumn{2}{l|}{\emph{$\beta$-skeleton Graph (2d)}}         & 113.27      & 3.20          & 35.37x           \\
\multicolumn{2}{l|}{\emph{Spanner (2d)}}        & 27.19       & 2.15          & 12.67x           \\ \hline
\end{tabular}
\caption{Runtimes (seconds) and parallel speedups ($T_1/T_{36h}$) for \pargeo{} implementations on uniform hypercube data sets of varying dimensions and 10 million points. $T_1$ and $T_{36h}$ denote the single-threaded and the 36-core hyper-threaded times, respectively. For batch-dynamic \kdtree{} updates, each batch contains $10\%$ of the data set.} \label{tab:all-bench}
\end{table}

%% file: sections/hull.tex
\section{Convex Hull}\label{sec:hull}

\input{sections/hull/figure-hull-seb}

The convex hull of a set of points $P$ in $\mathbb{R}^d$
is the smallest convex polyhedron containing $P$.
%(in this paper, we assume the dimensionality $d$ is a constant).
It is common to represent the convex hull using a set of
\defn{facets}.
The boundary of two facets is a \defn{ridge}.
For example, in $\mathbb{R}^3$,
assuming the points are in general position
(no four points are on the same plane),
each facet is a triangle, and each ridge is a line
that borders two facets (see Figure~\ref{fig:defs}(a)).

The randomized incremental algorithm and the quickhull algorithm
are the most widely used algorithms for solving convex hull in practice.
The randomized incremental algorithm for $\mathbb{R}^d$ was
proposed by Clarkson and Shor~\cite{clarkson89random}.
Given a point data set $P$ in $\mathbb{R}^d$,
the randomized incremental algorithm first constructs a $d$-simplex,
a generalization of a tetrahedron in $d$-dimensions as the initial hull.
Then, the algorithm adds the points to the polyhedron in a random order, updating the hull if necessary.
% The overall work of the algorithm is $O(n\log n + n^{\lfloor d/2 \rfloor})$ in expectation.
In practice, the quickhull algorithm~\cite{greenfield1990, barber1996},
another incremental algorithm, is often used. Unlike the randomized
algorithm, the quickhull algorithm processes a point that
is furthest from a facet, which enables the hull to be expanded more quickly.
% The key improvement of the quickhull algorithm is that, at each time step, rather than processing a random point, the algorithm processes a point that is furthest from a facet, which enables the hull to be expanded more quickly.
The quickhull algorithm is by far one of the most common implementations for convex hull due to its simplicity and efficiency~\cite{leomccormack-hull-impl,qhull-hull-impl,quickhull3d-hull-impl,akuukka-hull-impl,karimnaaji-hull-impl,cgal}.
There have also been works that study parallel implementations of quickhull, but they are either limited to $\mathbb{R}^2$~\cite{nher2008divide, srikanth2009parallelizing}, or do not return the exact convex hull for $\mathbb{R}^3$~\cite{stein2012, tzeng2012finding}.
Recently, Blelloch~\etal~\cite{blelloch2020} proposed a new randomized incremental algorithm that is highly parallel in theory. However, the algorithm does not seem to be practical due to numerous data structures required for bookkeeping.
% Sequential and parallel algorithms based on divide-and-conquer~\cite{preparata1977hull,aggarwal1988parallel,dadoun1989,amato1992parallel} are not known to be practical for $\mathbb{R}^3$ and above, due to the use of complicated subroutines to merge convex hulls. Other approaches, such as the gift-wrapping algorithm~\cite{chand1970gift}, Graham scan~\cite{graham1972}, and Jarvis march~\cite{jarvis1973}, are either not parallel or are limited to solving convex hull in $\mathbb{R}^2$.

%% End of prelim

In this section, we describe our new parallel reservation-based algorithm.
Our algorithm is able to express both the randomized
incremental convex hull algorithm and the quickhull algorithm.
Specifically, unlike a sequential incremental algorithm that adds one
point per round, we add multiple points in parallel per round.
We resolve conflicts caused by the parallel insertion using a
reservation technique.
%It can also serve as an effective building block
%for many of the GPU-CPU hybrid algorithms mentioned above
We also apply a general parallelization technique
based on divide-and-conquer, which in combination with our parallel
incremental algorithm, leads to faster implementations in practice.
% In addition, we proposed new optimizations for the convex hull and compare with existing ones in the multi-core context. \yiqiu{the optimization part might be optional}
% \yan{We may want to emphasize the ``reservation technique'' in this section, ideally in a subsection title, or at least in a paragraph title, since we claimed that it is the main algorithmic insight.}\yiqiu{Sounds good. Added reservation technique to the text above}

\myparagraph{Parallel Reservation-Based Algorithm}\label{sec:incremental}
Our parallel reservation-based algorithm can be implemented
as either a randomized incremental algorithm or a quickhull algorithm.
We will first introduce the overall structure of the algorithm.
Then, we will describe the details about the implementations,
and compare with existing approaches.
We will base our description in the context of
$\mathbb{R}^3$ for the sake of clarity, but the algorithm
can be extended to $\mathbb{R}^d$ for any constant integer $d\ge 2$.

We first give a high-level overview of the algorithm.
Given an ordered set of points $P=\{p_1,p_2,\dots,p_n\}$,
we let $P_r=\{p_1,p_2,\dots,p_r\}$ be the prefix of $P$ of size $r$, and $CH(P_r)$ be the convex hull on $P_r$.
We start the construction by first arbitrarily selecting four points from $P$ that do not lie on the same plane and putting them at the beginning of $P$, forming a tetrahedron $CH(P_4)$.
% We then make these four points the first four in $P$, and denote the tetrahedron as $CH(P_4)$.
% We randomly permute $P$, after which we pick the first four points (assuming they do not lie on the same plane), forming a tetrahedron $CH(P_4)$.
%\julian{what does it mean that "We then make these four points the first four in $P$"? should we just delete that phrase?}\yiqiu{Sounds good I deleted it.}
Then, the algorithm proceeds iteratively,
but on each round,
rather than inserting just $p_r$ to form $CH(P_r)$,
we process a batch of points in parallel.
On each round,
let each point outside of $CH(P_{r-1})$ be called a
\defn{visible point}.
We first select a batch of visible points,
and try to add them to $CH(P_{r-1})$ in parallel in
the same round.

The key challenge of this approach is that some of these
points cannot be processed in parallel due to
concurrent modifications on the shared structures of the
convex polyhedron.
We use a reservation algorithm to resolve these conflicts,
such that we only process the points that modify disjoint
facets of the polyhedron.
Specifically, each point will perform a priority write~\cite{Shun2013} with its ID to reserve all of its visible facets.
Points that have their ID written to all of its visible facets are \emph{successful}.
We then process the successful points in parallel
by enabling them to make modifications to
$CH(P_{r-1})$.
At the end of the round, in parallel, we filter out points that are no longer visible.
The algorithm will terminate when there are no more visible points.

\begin{figure}
\centering
\vspace{-5pt}
  \includegraphics[width=0.55\linewidth]{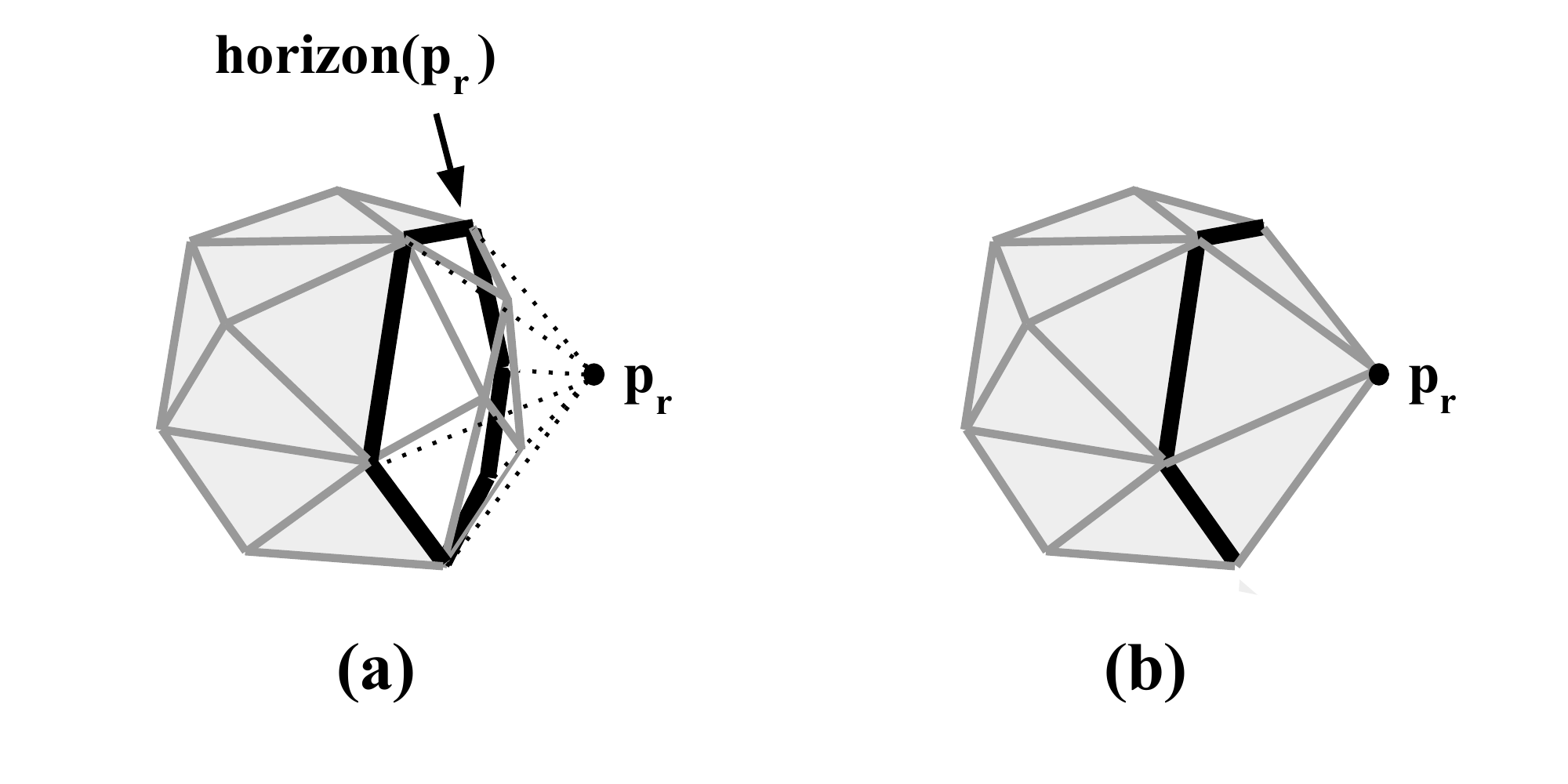}
  \vspace{-5pt}
  \caption{Illustration of adding a visible point $p_r$ to the convex hull. (a)
  shows the convex hull prior to the addition of $p_r$. The visible facets are in white,
  while the non-visible facets are in gray. The thicker line segments correspond to the
  horizon. (b) shows the convex hull after adding $p_r$ with newly created facets.
  }\label{fig:hull-inc}
\end{figure}

%%% \myparagraph{Detailed Algorithm}
We now describe the algorithm in more detail.
Figure~\ref{fig:hull-inc} illustrates the processing of a 
visible point $p_r$.
We denote a facet as a \defn{visible facet} of $p_r$ if point $p_r$ is in the half space away from the center of the convex hull. 
We first retrieve the set of visible facets of $p_r$ via facets stored
in it.
The visible facets of $p_r$ form a closed region,
whose boundary is a set of ridges, known as the \defn{horizon}.
We delete the visible facets from $CH(P_{r-1})$, and replace them with new facets,
where each new facet is formed by adding two ridges from
a horizon ridge to $p_r$.

\begin{figure}[t]
\centering
  \includegraphics[trim=0 50 0 50,clip,width=0.33\linewidth]{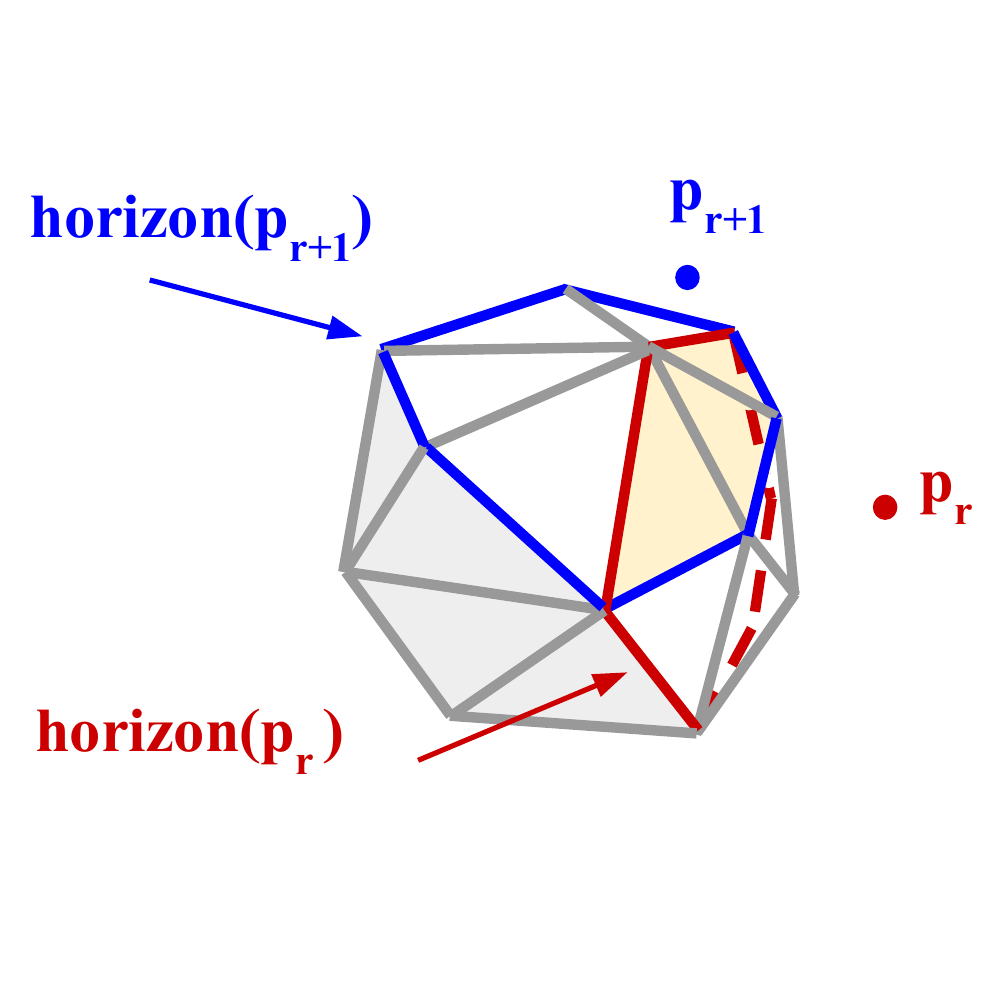}
  \caption{This figure illustrates the attempt to add $p_r$ and $p_{r+1}$ in parallel.
  The visible points and horizons of $p_r$ and $p_{r+1}$ are in red and blue,
  respectively. The visible facets to either visible points are in white/yellow,
  while the other facets are in gray. The overlap of the three visible facets between the
  $p_r$ and $p_{r+1}$ is in yellow. 
  }\label{fig:hull-conflict}
\end{figure}

Because of the structural changes to the convex hull that occur
when adding a visible point, concurrent structural changes
can cause data races, which need to be avoided.
We show an example of the conflict in Figure~\ref{fig:hull-conflict},
where we are attempting to add two visible points
$p_r$ and $p_{r+1}$ in parallel.
As shown in the figure, the closed regions formed by the visible
facets of each visible point overlap with each other in three facets, which are highlighted in yellow.
Should the two visible points be processed in parallel,
the resulting polyhedron may not be well-defined due to data races.
When processed sequentially, $p_{r+1}$'s visible facets would have
been different, involving newly created facets by $p_r$.
%Therefore, for this example, our parallel algorithm only allows
%one of $p_r$ and $p_{r+1}$ to be processed in the same round.

%Input: 3-dimensional points @$P=\{p_1,p_2,\dots,p_n\}$@, batch size @$r$@
\begin{figure}[t]
\begin{minipage}[t]{\columnwidth}
\begin{lstlisting}[linewidth=.99\columnwidth, numbers=left, xleftmargin=3.5ex]
Input: 3-dimensional points P, batch size r
Output: 3-dimensional convex hull
CH := initialize with 4 points @\label{line:hull-init}@
while (P is not empty): @\label{line:hull-while}@
  Q := a batch of size @$r$@ of visible points in P @\label{line:apex}@
  par_for (q in Q): /* reservation */ @\label{line:reserve-s}@
    for (f in q.visibleFacets):
      @\writeMin{}@(&f.reservation, q.id) @\label{line:reserve-e}@
  par_for (q in Q): /* check reservation */ @\label{line:check-s}@
    for (f in q.visibleFacets):
      q.success &&= (f.reservation == q.id) @\label{line:check-e}@
  par_for (q in Q): /* process successful points */ @\label{line:add-s}@
    if (q.success):
      @delete@ q's visible facets
      create new facets of q
      update CH @\label{line:add-e}@
  P := @\pack{}@(P, visible) @\label{line:pack}@
\end{lstlisting}
\end{minipage}\hspace{.3in}

\caption{Pseudocode for the parallel reservation-based convex hull algorithm
(which includes the randomized incremental algorithm and the quickhull algorithm).
}
\label{code:hull-high-level}
\end{figure}

Our reservation algorithm allows only a subset of the visible points
that update disjoint facets of the convex hull to be processed
in parallel on each round.
At a high level, we use the lexicographical order of the visible
points to determine the priority in processing a facet (a smaller ID has higher priority).
In the example shown in Figure~\ref{fig:hull-conflict},
since $p_r$ has a smaller ID
than $p_{r+1}$, the three conflicting facets can only be processed
by $p_r$ in that round.
The pseudocode for the algorithm is shown in Figure~\ref{code:hull-high-level}. 
$P$ is processed
iteratively until it is empty (Line 4).
We allocate an extra data field in each facet for performing reservations (Lines~\ref{line:reserve-s}--\ref{line:reserve-e}).
For each visible point in parallel, we iterate through its visible facets
and use a parallel priority write (\writeMin{}) to write its ID to the
facets' ``reservation'' fields.
% Note that for each facet, only the token of the visible point with the
% smallest lexicographical order will be kept.
Then on Lines~\ref{line:check-s}--\ref{line:check-e}, we determine
which visible points successfully reserved all of its facets.
Again, in parallel for each visible point, we check each of its
visible facets for a successful reservation by comparing the value of
the reservation field with its token.
The reservation of a visible point is only successful if its ID is stored in all of its visible facets.
Then, on  Lines~\ref{line:add-s}--\ref{line:add-e}, we process the visible points whose reservations are
successful, adding them to the hull and updating the appropriate data structures.
Finally, on Line~\ref{line:pack}, we process the points
in $P$ such that those remaining as visible points are
packed to replace the original $P$, and the non-visible points are discarded. 
Note that the visible points that succeeded in the reservation are
no longer visible points because they are now part of the convex hull.
Some of the remaining points will also no longer
be visible points due to the growth of the convex hull.

We use a simple and fast data structure to keep track of the
visibility relationship between the visible points and the facets.
At each step of the algorithm, when a visible point
is processed, it needs to identify the set of visible facets.
On the other hand, for the facets undergoing
structural changes, they need to identify and
redistribute their visible points to new facets.
To find the set of visible facets of $p_r$,
it is inefficient to iterate through all of the facets of $CH(P_{r-1})$.
While existing approaches~\cite{deberg2000compgeom}
keep track of the visibility between visible points and \emph{all} of their
visible facets, we found such an approach to be slow
becaseu each vertex is associated with
multiple facets, making the cost of storing and updating the data structure high.
We only store the reference of an arbitrary visible facet to each
visible point, from which we use a local breadth-first search to
retrieve all of the visible facets only when needed. 
For storing the visible points in the facets, we assign each point to one of its visible facets.
During point redistribution, we gather the points stored in each 
visible facet into an array, 
and in parallel distribute each point to a new visible facet replacing the original visible facet. Each such point also stores a reference to this visible facet.

Our reservation-based algorithm can be used to implement the parallel randomized
incremental algorithm or the quickhull algorithm for convex hull.
For the randomized incremental algorithm, we randomly permute the input points at the beginning, and on each round attempt to add a prefix
of the permuted points to the convex hull.
For the quickhull algorithm, on each round, we instead select a set of points furthest
from a subset of facets.
% We describe the reservation-based algorithm, and the implementation of the two algorithms in greater detail in Appendix~\ref{appendix:hull-detailed}.
\iffull
We describe the implementation of the two algorithms in greater detail in Appendix~\ref{appendix:hull-detailed}.
\else
We describe the implementation of the two algorithms in greater detail in the full version of our paper.
\fi
Our reservation-based algorithm is inspired by the idea of "deterministic reservations" from Blelloch \etal~\cite{BlellochFGS12}, who introduce this approach to implement parallel algorithms for other problems.
\iffull
We also show the work overhead of doing reservations compared to the sequential algorithm is small in Appendix~\ref{appendix:hull-overhead}.
\else
In the full version of our paper,
we show the work overhead of doing reservations compared to the sequential algorithm is small.
\fi

%% Detailed randomized and quickhull moved to appendix

%% Overhead of reservation moved to the appendix

%% Pseudo hull content moved to the appendix

\myparagraph{Parallel Divide-and-Conquer}\label{sec:dchull}
We adopt a common parallelization strategy using divide-and-conquer,
which calls our reservation-based algorithm as a subroutine.
Some  early convex hull algorithms are based on
divide-and-conquer, notably, the algorithm by Preparata and Hong~\cite{preparata1977hull}.
The algorithm splits the input into two spatially disjoint subsets
by a mid-point along one of the axis, recursively computes the convex hull on each subset, and then merges the results together.
Later work~\cite{aggarwal1988parallel,dadoun1989,amato1992parallel}
extended this approach to the parallel setting.
However, most of these approaches rely on complicated subroutines
to merge convex hulls, which are not practical and have not been implemented, to the best of our knowledge.

We implement a practical divide-and-conquer algorithm by partitioning
the input into $c\cdot numProc$ equal subsets, where $c$ is a small
constant and $numProc$ is the number of processors.
For each subset, the convex hull of the subset is computed by a single processor using
the sequential quickhull algorithm, but run in parallel across the different subsets.
Then, the vertices of the outputs of the subproblems are collected to
form a new input, from which the final convex hull is computed
using our reservation-based parallel algorithm described earlier.

% We also implement Tang~\etal's pseudohull algorithm for point culling~\cite{tang2012}, after which we use our reservation based algorithm to compute the true hull. The algorithm starts from an initial tetrahedra, and recursively grows each facet into three new facets using the furthest point from the facet. This results in a polyhedron, and the points in the interior of the polyhedron are pruned before the final computation. We include more details of this algorithm in Appendix~\ref{appendix:pseudo-hull}.

% \section{Point Culling via Pseudohull Computation} \label{appendix:pseudo-hull}

\myparagraph{Point Culling via Pseudohull Computation}
We also implement a multicore variant of Tang~\etal's pseudohull heuristic~\cite{tang2012}, originally proposed for the GPU.
Starting from an initial tetrahedra, we recursively grow each facet into three new facets, using the furthest point from the facet, similar to the quickhull algorithm. The visible points associated with the facet are redistributed to the new facets.
This results in a polyhedron, and the points in the interior of the polyhedron will not be part of the convex hull.
Therefore, we can prune away the points inside the polyhedron and compute the convex hull on the rest of the points.

There are several differences in our implementation from Tang~\etal's algorithm.
Our implementation executes the recursive calls on different facets asynchronously in parallel, 
whereas Tang~\etal's implementation maps the algorithm to the GPU architecture by
pre-allocating space for the facets and visible points, and runs the algorithm
in an iterative manner in lock-step.
Specifically, successively generated facets and points points associated them
are updated by multiple threads in parallel in each iteration.
We use a parallel maximum-finding routine to find the furthest point
of each facet in each call.
Rather than growing the pseudohull until there are no more visible points as done by Tang~\etal, we set a threshold on the number of points associated with a facet, below which we stop growing the pseudohull. This prevents stack overflow on large and skewed data sets 
%\julian{(to explain why this is not an issue for Tang's code)}\yiqiu{I think it's true, not all data sets have stack overflow. In the test it overflowed on skewed data set, but I suspect it's also possible on large enought data. On the non-overflowing data sets, setting the threshold higher sometimes has some speedup, but the advantage timing is noisy and the difference in threshold is minute, so maybe we don't need to mention it.} 
due to too many recursive calls, and the extra unpruned points do not contribute significantly to the work of the final computation of the convex hull.
At the end of pruning, we use our parallel reservation-based quickhull
algorithm to compute the final hull on the remaining points, whereas
Tang~\etal uses a sequential implementation.

%% file: sections/hull/figure-hull-seb.tex
\begin{figure}
\centering
  \includegraphics[width=0.5\linewidth]{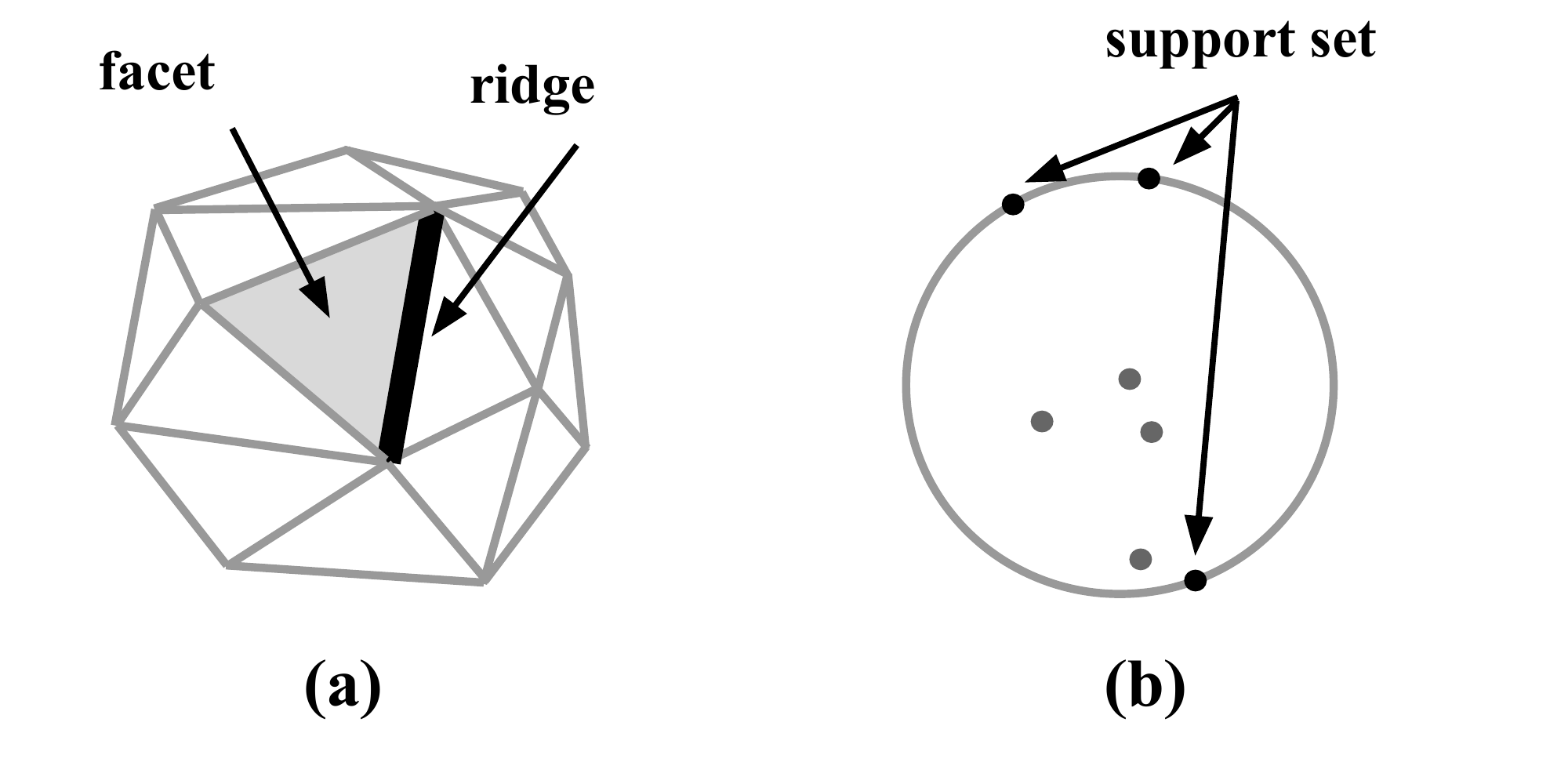}
  \caption{(a) A facet and a ridge of a convex hull in $\mathbb{R}^3$.
  (b) The support of the smallest enclosing ball in $\mathbb{R}^2$.
  }\label{fig:defs}
\end{figure}

%% file: sections/seb.tex
\section{Smallest Enclosing Ball}\label{sec:seb}
\label{sec:seb}

The smallest enclosing ball of $P$ in $\mathbb{R}^d$ is the smallest
$d$-sphere containing $P$.
It is well known that the smallest enclosing
ball is unique and defined by a \defn{support set} of $d+1$ points on the surface of the ball
(see Figure~\ref{fig:defs}(b)).

Welzl~\cite{welzl1991disks} showed that by using a randomized incremental algorithm,
the smallest enclosing ball can be computed in $O(n)$ time 
in expectation for constant $d$.
The algorithm iteratively expands the
support set of the ball by adding points in a random order until the ball contains all of the points.
The algorithm was later improved by Gartner~\cite{gartner1999seb}
with practical optimizations for speed and robustness.
Larsson~\etal~\cite{larsson2016} proposed practical parallel algorithms
that use a new method for expanding the support set, and their implementations work on both CPUs and GPUs.
Later, Blelloch~\etal~\cite{blelloch2016incremental} proposed a
parallel algorithm based on Welzl's algorithm, but without any
implementations.

% In this paper, we develop a parallel algorithm that improves upon Larsson~\etal~\cite{larsson2016}'s algorithm. We also provide the first implementation of Blelloch~\etal~\cite{blelloch2016incremental}'s algorithm.

%% End of prelim

In this section, we describe our new algorithms for the 
smallest enclosing ball problem based on Larsson~\etal's approach~\cite{larsson2016}.
We propose a sampling-based algorithm to quickly reduce
the size of the data set.
We also provide the first parallel implementation of Welzl's
classic algorithm.

%We also implement the parallel algorithm by
%Blelloch~\etal~\cite{blelloch2016incremental} for comparison.
%We show that our algorithm demonstrates significant running time improvement in the practice.

% \myparagraph{Sampling-Based Algorithm}\label{sec:seb-sampling}
Given a ball $B$, we define \defn{visible points} to be points that lie outside of $B$. Existing approaches for computing the smallest enclosing ball focus on expanding the support set in an iterative manner, and output the enclosing ball when there are no more visible points. Welzl's algorithm expands the support set by adding points in a random order~\cite{welzl1991disks}. In comparison, Larsson~\etal's approach scans the input to search for  good support sets in a round-based manner. In $\mathbb{R}^3$, Larsson's algorithm divides the space into eight orthants centered at the center of $B$. On each round, the input is scanned to find the furthest visible points in each orthant. $B$ is then updated to the next intermediate solution using the existing support set of $B$ and the new visible points found during the scan. The algorithm iterates until there are no more visible points.  It is parallelized within each round by performing the scan on the input in parallel.

\myparagraph{Sampling-Based Algorithm}\label{sec:seb-sampling}
We find each iteration in Larsson~\etal's algorithm to be unnecessarily expensive due to having to scan the entire data set on every round.
Our approach is to use a sampling heuristic to first obtain a good initial 
%\julian{obtain which support set? do you just mean you are solving the SEB problem?}\yiqiu{Yes, there's an example of the support set in Figure~\ref{fig:defs}b. Basically the sampling is to look at a small part of the data set but obtain a good initial ball} 
ball, inspired by Welzl's randomized algorithm.
Specifically, we use small random samples to obtain good estimates
of the support set at a negligible cost.

We show the pseudocode of our algorithm in Figure~\ref{code:seb-sampling}.
Our sampling-based algorithm consists of two phases: the sampling
phase (Line~\ref{line:seb-sample-s}--\ref{line:seb-sample-e})
and the final compute phase (Line~\ref{line:seb-final-s}--\ref{line:seb-final-e}).
First, we initialize the ball using a few arbitrary points (Line~\ref{line:seb-init}). 
Then, we iterate through a random permutation of the input to take multiple samples (Line~\ref{line:seb-sample-s}--\ref{line:seb-sample-e}).
On each iteration, we scan through a constant-sized segment of the unseen part of the input,
which is equivalent to a random sample.
We perform an orthant scan similar to Larsson's approach.
Our implementation of orthant scan will return a new estimate of
the support set based on the sample, and a boolean \textit{hasOutlier} indicating
whether the sample contains visible points with respect to the current smallest
enclosing ball $B$ (Line~\ref{line:seb-sample-scan}).
We recompute $B$ using the new support set.
If there are visible points in the current sample, we will continue the sampling process with our new $B$.
If there are no visible points in the sample,
the support set likely contains most of the points, and
so we terminate sampling and move on to the next phase.
Now, with a good estimate of the optimal smallest enclosing ball, we run Larsson's orthant scan to compute the final smallest
enclosing ball (Line~\ref{line:seb-final-s}--\ref{line:seb-final-e}). The sampling phase allows us to generate good support sets without having to scan the entire input.

%\julian{from the pseudocode, it may not be clear to the reader what the difference between the two phases is. they look almost exactly the same, except for the first argument to orthantScan. It would be useful to emphasize the difference here. Also the first argument to orthantScan P[scanned+c] should not be a single point but rather a range of points, probably $P[scanned,\ldots,\min(scanned+c,n)-1]$. c should also be an input to the algorithm.}\yiqiu{Fixed the pseudocode}\yiqiu{todo, point out difference int the two phase more clearly}

\begin{figure}
\begin{minipage}[t]{\columnwidth}
\begin{lstlisting}[linewidth=.99\columnwidth, numbers=left, xleftmargin=3.5ex]
Input: d-dimensional points P, batch size c
Output: d-dimensional smallest enclosing ball
B = ball() @\label{line:seb-init}@
/* Sampling phase */
scanned = 0 @\label{line:seb-sample-s}@
while (scanned < n):
  hasOutlier, support = @\label{line:seb-sample-scan}@
    orthantScan(P[scanned:min(scanned+c,n)-1],B) 
  scanned += c
  if (!hasOutlier):
  	break /* current sample does not violate B */
  else
  	B = constructBall(support) @\label{line:seb-sample-e}@
/* Final computation phase */
while (hasOutlier): @\label{line:seb-final-s}@
  hasOutlier, support = orthantScan(P, B)
  if (!hasOutlier):
  	return B
  else
  	B = constructBall(support) @\label{line:seb-final-e}@
\end{lstlisting}
\end{minipage}\hspace{.3in}

\caption{Pseudocode for the parallel sampling-based algorithm for smallest enclosing ball.
}
\label{code:seb-sampling}
\end{figure}

We parallelize the orthant scan, which is the
most expensive operation of the algorithm.
Specifically, we divide the input array to orthant scan into
blocks, and process each block sequentially, but in parallel across different blocks.
Afterward, the extrema for the orthants obtained from the
blocks are merged, and a new support set is computed on these points and the existing support set of $B$.

We parallelize the orthant scan, which is the
most expensive operation of the algorithm.
Specifically, we divide the input array to orthant scan into
blocks, and process each block sequentially, but in parallel across different blocks.
Afterward, the extrema for the orthants obtained from the
blocks are merged, and a new support set is computed on these points and the existing support set of $B$.

% We also implemented and optimized the parallel version of Welzl's algorithm described by Blelloch~\etal~\cite{blelloch2016incremental}.  The algorithm takes a random permutation of the input, and processes exponentially growing prefixes in parallel. When encountering a visible point outside of the current ball, a new ball is computed by recursively calling the parallel algorithm. We implement the parallel algorithm with the move-to-front heuristic~\cite{welzl1991disks}, where a visible point, when identified, will be moved to the front of the sequence, so it will be processed earlier in recursive calls. We also parallelize the pivoting heuristic by Gartner~\cite{gartner1999seb}. In this heuristic, upon encountering a visible point, rather than processing the visible point directly,
% we process a \emph{pivot point} furthest away from the center of the current ball.
% \iffull
% We describe the algorithm and optimizations in Appendix~\ref{appendix:seb-welzl} in greater detail. 
% \else
% We describe the algorithm and optimizations in the full version of the paper in greater detail. 
% \fi
\myparagraph{Parallel Welzl's Algorithm and Optimizations}
We also implemented and optimized the parallel version of Welzl's algorithm described
by Blelloch~\etal~\cite{blelloch2016incremental}.
Welzl's sequential algorithm uses a random permutation of the input $P$
and processes the points one by one.
If the algorithm encounters a visible point $p_i$ with respect
to the current bounding ball $B$, $B$ is
recomputed on $P_i$, the prefix of points up until $p_i$,
using recursive calls to the algorithm.
Blelloch~\etal's parallel algorithm also uses a random
permutation of $P$. Across iterations,
the algorithm processes prefixes of $P$ of exponentially increasing size.
If the prefix contains at least one visible point,
the earliest visible point $p_i$ is identified, and $B$ is recomputed on
prefix $P_i$ by recursively calling the parallel algorithm.
Each prefix is processed in parallel.

We implement the algorithm with some practical optimizations.
When there are numerous visible points in the prefix, the work of the parallel
algorithm will increase significantly, because each time a visible point is discovered,
the points after the visible point in the same prefix will have to be reprocessed in the next round.
Therefore, given that there will be more visible points in the initial rounds when the prefix size is small ($<500000$),
we process these prefixes sequentially by calling Welzl's sequential algorithm.
% \laxman{can we give some idea of what small is, e.g., prefixes of size smaller than a few thousand / tens of thousand?}\yiqiu{Sounds good, fixed}
This also reduces the amount of overhead from parallel primitives, since there is limited
parallelism for small prefixes.

In addition, we extend existing optimizations of Welzl's sequential algorithm to the parallel setting.
We implement the move-to-front heuristic~\cite{welzl1991disks},
which upon encountering a visible point, moves the visible point to the front of $P$,
so that it will be processed earlier in recursive calls, reducing the number of subsequent visible points.
% The parallel version of this heuristic is similar. However, because it is difficult to process prefixes of a linked list in parallel, we use an array to represent $P$ and
% simply swap the visible point with the point at the front of the array.
We also parallelize the pivoting heuristic proposed by Gartner~\cite{gartner1999seb}. 
In this heuristic, upon encountering a visible point, rather than processing the visible point directly,
we search $P$ for a \emph{pivot point} furthest away from the center of the current
$B$, and use the pivot point to compute the new $B$ instead of the visible point. 
We use a parallel maximum-finding algorithm to identify the pivot point.

%% file: sections/batch-kdtree.tex
\input{batch-kdtree/sections/algorithm}

% \input{batch-kdtree/sections/implementations}

%% file: batch-kdtree/sections/algorithm.tex
% \section{\ourtree{}} \label{sec:ourtree}
\section{Parallel Batch-Dynamic \kdtree{}} \label{sec:ourtree}

% \input{batch-kdtree/sections/introduction/fig-log-method}

% The concept of a parallel batch-dynamic data structure has become popular in recent years~\cite{WangY0S21,dhulipala2021parallel} as an important paradigm due to the availability of large (dynamic) data sets undergoing rapid changes. The idea is to batch together operations of a single type and perform them as a single batched update, rather than one at a time.

The \kdtree{}, first proposed by Bentley~\cite{bentley1975}, is a binary tree data structure that arranges and holds spatial data to speed up spatial queries. At each node, the data set is split  into two using an axis-aligned hyperplane along a dimension, until the node holds a small constant number of points. \kdtrees{} are used in a wide range of applications, such as in databases, machine learning, data compression, and cluster analysis.
% Given a set $P$ of $n$ $d$-dimensional points, the \kdtree{} is a balanced binary tree where each node represents a bounding box of a subset of the input points. The root node represents all of the points (and thus the tightest bounding box that includes all the points in $P$). Each non-leaf node holds a splitting dimension and splitting value that splits its bounding box into two halves using an axis-aligned hyperplane in the splitting dimension. Each child node represents the points in one of the two halves.
% This recursive splitting stops when the nodes hold some small constant number of points---these nodes are the leaves and directly represent the points.

In this section, we introduce the \ourtree{}, a parallel batch-dynamic \kdtree{} implemented using the logarithmic method~\cite{bentley-logarithmic-1, bentley-logarithmic-2}.
% (discussed in Section~\ref{prelim-log-method}). 
Our \ourtree{}s build on ideas from the Bkd-Tree by Procopiuc \etal~\cite{bkd} and the cache-oblivious \kdtree{} by Agarwal \etal~\cite{pankaj-co}.
% The structure is depicted in Figure~\ref{fig:logmethod}.
The logarithmic method~\cite{bentley-logarithmic-1, bentley-logarithmic-2} for converting static data structures into dynamic ones is a very general idea. At a high level, the idea is to partition the static data structure into multiple structures with exponentially growing sizes (powers of 2). Then, inserts are performed by only rebuilding the smallest structure necessary to account for the new points.
In the specific case of the \kdtree{}, a set of $N_s$ static \kdtrees{} is allocated, with capacities $[2^{0}, 2^{1},\ldots, 2^{N_s-1}]$, as well as an extra buffer tree with size $2^{0}$. Then, when an insert is performed, the insert cascades up from the buffer tree, rebuilding into the first empty tree with all the points from the lower trees.
If desired, the sizes of all of the trees can be multiplied by a buffer size $X$, which is a constant that is tuned for performance.
% In Figure~\ref{fig:logmethod}, all of the trees shown are full; one can imagine that the tree with size $2^{3}X$ is empty, so the next insert would cause the buffer and trees 0, 1, and 2 to cascade up to it.

We implement the underlying static \kdtrees{} in an \ourtree{} using the van Emde Boas (vEB)~\cite{arge-cache-oblivious-book,demaine2002cache,pankaj-co} recursive layout. 
Agarwal \etal~\cite{pankaj-co} show that this memory layout can be used with \kdtrees{} to make traversal cache-oblivious.
\iffull
We provide more details of the static tree structure, and parallel algorithms for the construction, deletion, and \knn{} search in Appendix~\ref{section:single-alg-top}.
\else
We provide more details of the static tree structure, and parallel algorithms for the construction, deletion, and \knn{} search in the full version of our paper.
\fi
%, although dynamic updates on a single tree become very complex. 
%However, in the logarithmic method, the underlying \kdtrees{} themselves are static, and so we are able to sidestep the complexity of cache-oblivious updates on these trees and benefit from the improved cache performance of the vEB layout. 
% For the buffer region of the \ourtree{}, we use a regular \kdtree{}, laid out like a binary-heap in memory (i.e., nodes are in a contiguous array, and the children of index $i$ are $2i$, $2i+1$).
% We will discuss the key parallel algorithms that we used in our implementation: construction, deletion, and \knn{} on the underlying individual \kdtrees{} (Section~\ref{section:single-alg-top}) and construction, insertion, deletion, and \knn{} on \ourtree{} (Section~\ref{section:log-alg-top}). Note that we do not need to support insertions on individual \kdtrees{}, because our \ourtree{} simply rebuilds the necessary \kdtrees{} upon insertions.
% We use subscript \textsc{S} to denote algorithms on the underlying \kdtrees{}, and subscript \textsc{L} to denote algorithms on the full logarithmic data structure.

% \iffull
% \else
% Due to space constraints, we defer the proofs of our theorem statements to the full version of the paper~\cite{full-paper-url}.
% \fi

\myparagraph{Parallel Batch Insertion}
Batch insertions are performed in the style of the logarithmic method~\cite{bentley-logarithmic-1, bentley-logarithmic-2}, with the goal of maintaining the minimum number of full trees within \logtree{}. Thus, upon inserting a batch $P$ of points, we rebuild larger trees if it is possible using the existing points and the newly inserted batch.
% This is implemented as shown in Algorithm~\ref{alg:log-insert}, and depicted in Figure~\ref{fig:bdl-insert}.
% This is implemented as depicted in Figure~\ref{fig:bdl-insert}.
We use a bitmask to determine which static trees in the structure to destroy and reconstruct
after each insertion.
Specifically, we build a bitmask $F$ of the current set of full static trees.
Given the buffer \kdtree{} size $X$, we add the value $\lfloor |P|/X \rfloor$ to $F$ when a point set $P$ is inserted, %\shangdi{the buffer tree has not been introduced}\yiqiu{It was accidentally deleted, I added it above when introducing the logarithmic method} 
% \shangdi{we can add $(|P|-|Q|)/X$ to $F$ when a point set $P$ is inserted and a subset of points $Q \subset P$ is added to the buffer tree}
after which the bitwise difference with the previous $F$ indicates which trees need to be changed. We gather the points in the trees to be destroyed, and with $P$, we construct
a subset of new trees in parallel.
As an implementation detail, note that we first add $|P| \mod{X}$ points to the buffer \kdtree{}---if we fill up the buffer \kdtree{}, then we gather the $X$ points from it and treat them as part of $P$, effectively increasing the size of $P$ by $X$.
% First, on line 2, we build a bitmask $F$ of the current set of full static trees in the logarithmic structure. Then, on line 3, because the buffer \kdtree{} has size $X$, we can add $|P|/X$ to $F$ to compute a new bitmask $F_{new}$ of full trees that would result if we added $|P|$ points to the tree structure. As an implementation detail, note that we first add $|P| \mod{X}$ points to the buffer \kdtree{}---if we fill up the buffer \kdtree{}, then we gather the $X$ points from it and treat them as part of $P$, effectively increasing the size of $P$ by $X$. Then, on line 4, taking the bitwise difference between these two bitmasks gives the set of trees that should be consolidated into new larger trees---specifically, any tree that is set in $F_{new}$ but not in $F$ must be constructed from trees that are set in $F$ but not in $F_{new}$. After determining which trees should be combined into new trees, on line 5 we construct all the new trees in parallel---in parallel for each new tree to be constructed, we deconstruct and gather all the points from trees that are being combined into it and then we construct the new tree over these points and any additional required points from $P$ using Algorithm~\ref{alg:veb-construct}.
Refer to Figure~\ref{fig:bdl-insert} for an example of this insertion method ($X>2$ in this example). In Figure~\ref{fig:bdl-insert-0}, the \ourtree{} contains $X$ points, giving a bitmask of $F=1$ (because only the smallest tree is in use). If we insert $X+1$ points, then we put one point in the buffer tree and compute $F_{new} = 1 + \lfloor \frac{X}{X} \rfloor = 2$, and so we have to deconstruct static tree 0 and build static tree 1, as shown in Figure~\ref{fig:bdl-insert-1}. Then, if we insert $X+1$ points again, then we again put one point in the buffer tree and compute $F_{new}=2 + \lfloor \frac{X}{X} \rfloor = 3$, and so we simply construct tree 0 on the $X$ new points (leaving tree 1 intact), as seen in Figure~\ref{fig:bdl-insert-2}. Finally, if we then insert $X-1$ points, this would fill the buffer up, and so we take 1 point from the buffer and insert $X$ points; then, $F_{new} = 3 + \lfloor \frac{X}{X} \rfloor = 4$, and so we deconstruct trees 0 and 1, and construct tree 2, as seen in Figure~\ref{fig:bdl-insert-3}.
\iffull
We include a more detailed explanation of the algorithm in Appendix~\ref{appendix:bdl-insert},
and explain the batch deletion algorithm in Appendix~\ref{sec:bdl-parallel-delete}.
\else
We include a more detailed explanation of the algorithm, and explain the batch deletion algorithm in the full version of our paper.
\fi

\input{batch-kdtree/sections/algorithm/fig-insert-example}

% \subsection{Data-Parallel $k$-NN}\label{alg:dpknn}
\myparagraph{Data-Parallel $k$-NN}
% \input{batch-kdtree/pseudocode/log-knn}
% In the data-parallel \knn{} implementation, we parallelize over the set of points given to search for nearest neighbors. First, on line 2, we allocate a \knn{} buffer for each of the points in $S$. Then, for each of the non-empty trees in \logtree{}, we call the data-parallel \knn{} subroutine on the individual tree, passing in the set $S$ of points and the set of \knn{} buffers. Because we reuse the same set of \knn{} buffers for each underlying \knn{} call, we eventually end up with the $k$-nearest neighbors across all of the individual trees for each point in $S$.
In the data-parallel \knn{} implementation, we parallelize over $S$, the set of points to search for nearest neighbors for. First, we allocate a \knn{} buffer for each of the points in $S$. Then, iterating over each of the non-empty trees in the \logtree{} sequentially, we call the data-parallel \knn{} subroutine on the tree, passing in the set $S$ of points and the \knn{} buffers. Because we reuse the same set of \knn{} buffers for each \knn{} call (note that each \knn{} call is internally parallel), we end up with the $k$-nearest neighbors of the entire pointset for each point in $S$.
\iffull
We include a more detailed explanation in Appendix~\ref{alg:dpknn}.
\else
We include a more detailed explanation in the full version of our paper.
\fi

% \input{batch-kdtree/sections/algorithm/single-tree}

% \input{batch-kdtree/sections/algorithm/parallel-deletion}

% \input{batch-kdtree/sections/algorithm/knn}

%% Detailed text moved to appendix
% \input{batch-kdtree/sections/algorithm/batch-dynamic}

%% file: batch-kdtree/sections/algorithm/fig-insert-example.tex
\begin{figure}[!t]
\centering
    \begin{subfigure}{0.23\textwidth}
        \centering
        \includegraphics[width=0.8\textwidth]{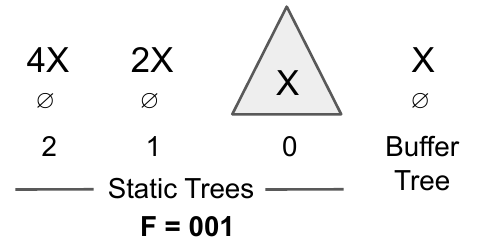}
        \caption{Static tree 0 is full.}
        \label{fig:bdl-insert-0}
    \end{subfigure}
    \hfill
    \begin{subfigure}{0.23\textwidth}
        \centering
        \includegraphics[width=0.8\textwidth]{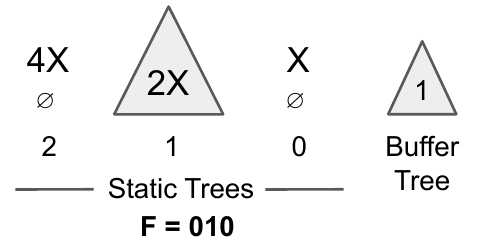}
        \caption{Static tree 1 is full and buffer tree has 1 point.}
        \label{fig:bdl-insert-1}
    \end{subfigure}
    \hfill
    \begin{subfigure}{0.23\textwidth}
        \centering
        \includegraphics[width=0.8\textwidth]{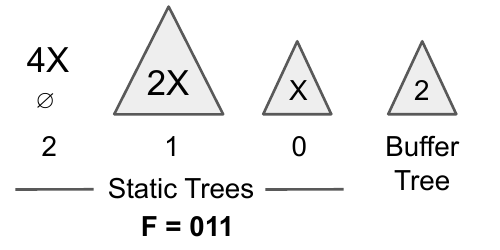}
        \caption{Static trees 0 and 1 are full and buffer tree has 2 points.}
        \label{fig:bdl-insert-2}
    \end{subfigure}
    \hfill
    \begin{subfigure}{0.23\textwidth}
        \centering
        \includegraphics[width=0.8\textwidth]{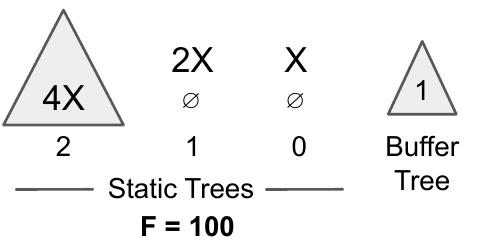}
        \caption{Static tree 2 is full and buffer tree has 1 point.}
        \label{fig:bdl-insert-3}
    \end{subfigure}
    \caption{A \ourtree{} in various configurations with $X>2$; starting from~(a), inserting $X+1$ points gives~(b), then inserting $X+1$ points gives~(c), and then inserting $X-1$ points gives~(d). %\shangdi{the reference to this figure is currently referencing the duplicated figure 14}\yiqiu{thanks, fixed}
    }
    \label{fig:bdl-insert}
\end{figure}

%% file: sections/experiments.tex
\section{Experimental Evaluation}
\label{sec:exp}

\myparagraph{Data Sets}
% Our synthetic data sets \defn{inSphere} and \defn{inCube} contain points generated uniformly at random in a sphere and a cube, respectively.
% We generate these data sets with size ranging from one thousand to one billion, in 2 and 3 dimensions. We name the data sets in the format of \defn{name\_dimension\_size}.
We use several types of synthetic data sets. The first is \textbf{Uniform} (\textbf{U}), consisting of points distributed uniformly at random inside a hypercube with side length $\sqrt{n}$, where $n$ is the number of points.
The second type \textbf{InSphere} (\textbf{IS}) is similar to the first, but the points are distributed in a hypersphere instead.
We also use \defn{OnSphere} (\textbf{OS}) and \defn{OnCube} (\textbf{OC}) data sets, where points are uniformly distributed on the surface of a hypersphere and a hypercube, respectively.
The surfaces have a thickness equal to 0.1 times the diameter or side length
of the sphere or cube.
% The last type is \textbf{VisualVar} (\textbf{V}), a clustered data set with variable density, produced by Gan and Tao's generator~\cite{gantao2015}. The generator produces points by performing a random walk in a local region, but jumping to random locations with some probability. For each of these two types, we generate them in 2, 3, 5, and 7 dimensions, and for 10 million points.
We name the data sets in the format of \textbf{Dimension-Name-Size}.

We also use the following real-world data sets from the Stanford 3D Scanning Repository~\cite{stanford2012}:
% \defn{thaiStatue\_3d\_5m} is a 3-dimensional point data set of size $4,999,996$ from a scanned statue; and
% \defn{asianDragon\_3d\_3.6m} is a 3-dimensional point data set of size $3,609,600$ from a scanned statue of a dragon.
\defn{3D-Thai-5M} is a 3-dimensional point data set of size $4999996$ from a scanned thai-statue; and
\defn{3D-Dragon-3.6M} is a 3-dimensional point data set of size $3609600$ from a scanned statue of a dragon.
% We also use 3 additional real-world data sets: \textbf{10D-H-1M}~\cite{ht-data set, ht-paper} is a 10-dimensional data set consisting of 928,991 points of home sensor data; \textbf{16D-C-4M}~\cite{chem-data set, chem-paper} is a 16-dimensional data set consisting of 4,208,261 points of chemical sensor data; and \textbf{3D-C-321M}~\cite{cosmo} is a 3-dimensional data set consisting of 321,065,547 points of astronomy data.  Due to time constraints, we only ran experiments on \textbf{3D-C-321M} using \ourtree{} in parallel to demonstrate that \ourtree{} can scale to large data sets.

\myparagraph{Testing Environment}
%We perform experiments on an Amazon EC2 instance with 2 $\times$ Intel Xeon Platinum 8124CL (3.00GHz) CPUs for a total of 36 cores with two-way hyper-threading, and 144 GB of RAM.  By default, we use all cores with hyper-threading.  We use the \texttt{g++} compiler (version 9.3.0) with \texttt{-O3} flag.
The experiments are run on an AWS c5.18xlarge instance with 2 Intel Xeon Platinum 8124M CPUs (3.00 GHz), for a total of 36 two-way hyper-threaded cores and 144 GB RAM. 
%Our experiments use all hyper-threads unless specified otherwise. 
We compile our benchmarks with the \texttt{g++} compiler (version 9.3.0) with the \texttt{-O3} flag, and use ParlayLib~\cite{parlaylib} for parallelism.
% All reported running times are the medians of 3 runs, after one extra warm-up run for each experiment.

% \myparagraph{Implementations}
% In our parallel implementations, we use the parallel primitives and the work-stealing scheduler implemented using \texttt{pthreads} in ParlayLib~\cite{parlaylib}.

\subsection{Convex Hull}

We test the following implementations for convex hull (our new implementations are underlined). All implementations are for both $\mathbb{R}^2$ and $\mathbb{R}^3$.%, unless stated otherwise \julian{i don't think we stated otherwise}.
\begin{itemize}[topsep=1pt,itemsep=0pt,parsep=0pt,leftmargin=10pt]
\item \defn{CGAL}: sequential C++ implementation of quickhull in CGAL~\cite{cgal}.% \shangdi{which algorithm is used in CGAL?}\yiqiu{Quickhull, fixed}.
\item \defn{Qhull}: sequential C++ implementation of quickhull~\cite{qhull-hull-impl} by
Barber~\etal~\cite{barber1996}.
\item \underline{\defn{RandInc}}: our implementation of the parallel
randomized incremental algorithm described in Section~\ref{sec:incremental}.
\item \underline{\defn{QuickHull}}: for $\mathbb{R}^2$, it is a simple recursive
parallel algorithm~\cite{blelloch1990vector}, and we use the
implementation in PBBS~\cite{pbbs};
for $\mathbb{R}^3$, we use our parallel quickhull algorithm described in Section~\ref{sec:incremental}.
\item \underline{\defn{Pseudo}}: our implementation of the pseudoHull heuristic proposed by Tang~\etal~\cite{tang2012} for 3-dimensional convex hull described in Section~\ref{sec:dchull}.
%Appendix~\ref{appendix:pseudo-hull}.
The final stage of the computation uses our \textit{quickHull} algorithm for $\mathbb{R}^3$.
% \shangdi{does this also work for $\mathbb{R}^2$? }\yiqiu{R2 uses a different implementation which is much simpler}\shangdi{maybe we should have a sentence about why this method is not in the plot for $R^2$? also just curious, how is the performance of the simpler pseudohull for R2 compared with other methods?}\yiqiu{Oh it's actually in there. It's in Figure 4. The Pseudo method only applies to R3, so we did not mention R2 here.}\shangdi{I see. Should we say that it only applies to R3? and why it only applies to R3? I suppose in R2 you can also replace an edge with two edges, but maybe this is not helpful in 2D? }\yiqiu{Sounds good, mentioned above, it works for 3d. For R2 if we do that it will give the correct hull directly, and no final computation is needed.}
\item \underline{\defn{DivideConquer}}: our divide-and-conquer algorithm described in Section~\ref{sec:dchull}.
\end{itemize}

\input{plots/hull2d}
\input{plots/hull3d}

In Figures~\ref{plot:hull2d} and~\ref{plot:hull3d},
We show a comparison of running times across different methods using 36 cores with two-way hyper-threading.
Our implementations achieve significant speedup compared
to existing sequential implementations.
Our fastest parallel implementations
achieve speedups of 190--559x (325x on average) over \textit{CGAL} for 2-dimensional convex hull,
and speedups of 10.5--124x (61.4x on average) over \textit{CGAL} for 3-dimensional convex hull.
Our fastest parallel implementations
have speedups of 147--1673x (605x on average) over 2-dimensional \textit{Qhull},
and speedups of 5.68--43.8x (19.9x on average) over 3-dimensional
\textit{Qhull}.
When run using a single thread, our parallel implementations achieve speedups of
3.26--12.4x and 1.31--5.05x over \textit{CGAL} for
2 and 3 dimensions, respectively;
and 3.39--47.6x and 0.99--2.06x speedups over \textit{Qhull}
for 2 and 3 dimensions, respectively. 
% \laxman{For the previous few sentences of results, can we provide some ideas for the reader about why our results are so much faster compared with cgal / quickhull? why are the speedups over 3d qhull worse than our speedups for 2d qhull?}\yiqiu{The main reason lies in the implementation. It's quite hard to know exactly at the moment, but I will try to figure something out}

For $\mathbb{R}^2$, \textit{DivideConquer} is always the fastest
method due to having high scalability from processing many independent subproblems in parallel.
For $\mathbb{R}^3$, the fastest two methods are \textit{DivideConquer}
and \textit{Pseudo}.
We observe that on data sets with a larger output size, \textit{Pseudo}
is slower than \textit{DivideConquer} (Figures~\ref{plot:hull3d}(a), (b), and (g)).
This is because the final computation after pruning takes longer
given that there are a higher number of remaining points after pruning.
For instance, the number of remaining points for 
\textit{3D-IS-10M} and \textit{3D-U-10M}  after pruning are  $83669$ and $2316$, respectively,
and \textit{Pseudo} is relatively slower on the former.
We observe that
\textit{RandInc} and \textit{QuickHull} take relative longer
compared with the fastest methods for
data sets with a smaller output size (Figures~\ref{plot:hull3d}(c)--(e) and (h)).
This is caused by higher contention during the reservation of facets,
since there are fewer facets on the intermediate hull.
For example, for \textit{3D-IS-10M} and \textit{3D-U-10M}, the output
sizes are $14163$ and $423$, respectively. During the
computation, \textit{3D-U-10M} exposes fewer facets for reservation,
leading to a lower success rate during the reservations.
% \input{plots/hull-speedup}

% Our parallel implementations achieve high self-relative speedups, as summarized in Table~\ref{table:hull-speedup}.
\textit{DivideConquer} achieves the best parallel
speedup (42.78x and 16.55x on average for $\mathbb{R}^2$
and $\mathbb{R}^3$, respectively). This is because the
bulk of the time is spent in computing independent convex hulls across different threads.
On the other hand, the incremental algorithms, \textit{RandInc}
and \textit{QuickHull}, demonstrate lower scalability because of load imbalance caused by the different amounts of work for each conflict point being processed in parallel.

% \input{plots/hull-vsn}

% We demonstrate this experiment and \emph{QuickHull}'s scaling versus the input
% size in Appendix~\ref{appendix:hull-seb-vsn}.
% Our implementations can process data sets with up to one billion points,
% which is 10x larger than the largest data set used in testing existing
% implementations for parallel architectures~\cite{stein2012,tang2012,gao2013,srikanth2009parallelizing,masnadi2020}.

%\julian{we cannot say this anymore since we don't have experiments on 1B points. can you add the experiments to the appendix (or maybe we can fit it here after removing comments)?}\yiqiu{It's strange to find it here because it looks like the largest we have used is just 321M. Maybe we can just remove this sentence from both papers.}
%\julian{in Figure 9 of the PPoPP submission we tested in a data set with 1 billion points}\yiqiu{Could you share that version of the paper? The ones in my overleaf and on arxiv did not have Figure 9.}\julian{https://www.overleaf.com/project/60f1a8d0af3de548d59f49dc}\yiqiu{Oh sorry I thought it's the bkdtree paper. The 1 billion point experiment is now added.}
% In Figure~\ref{plot:hull-vsn}, we plot the running time of our parallel quickhull implementation against the number of data points, ranging from one thousand to one billion. We observe a close-to-linear scaling as the data size increases.

\subsection{Smallest Enclosing Ball}
% \yiqiu{Try to include an analysis for why some implementation in the SEB are faster}

% \input{plots/seb-speedup}
% \input{plots/seb-vsn}

We test the following implementations for smallest enclosing ball (our new implementations are underlined). All implementations work for both $\mathbb{R}^2$ and $\mathbb{R}^3$.
\begin{itemize}[topsep=1pt,itemsep=0pt,parsep=0pt,leftmargin=10pt]
\item \defn{CGAL}: sequential C++ implementation of Welzl's algorithm in CGAL~\cite{cgal}.
\item \underline{\defn{Orthant-scan}}: our implementation of Larsson~\etal's orthant-scan algorithm~\cite{larsson2016}.
\item \underline{\defn{Sampling}}: our parallel sampling algorithm described in
Section~\ref{sec:seb}. 
\item \underline{\defn{Welzl}}: our parallel implementation of Welzl's
algorithm described in Section~\ref{sec:seb}. 
\item \underline{\defn{WelzlMtf}}: the same as \textit{Welzl}, but with the move-to-front heuristic~\cite{aggarwal1988parallel}.
\item \underline{\defn{WelzlMtfPivot}}: the same as \textit{Welzl}, but with both the move-to-front and the pivoting heuristic~\cite{gartner1999seb}.
\end{itemize}

For smallest enclosing ball,
we show the comparison across implementations using 36 cores with two-way hyper-threading in 
Figure~\ref{plot:seb}.
Our fastest parallel implementations
have speedups of 70--178x (109x on average) over \textit{CGAL}.
On one thread, our fastest implementations achieve speedup of 2.81--7.05x
(4.96x on average) over \textit{CGAL}.

\input{plots/seb}

Our sampling-based method is the fastest for eight out of the twelve
data sets,
whereas \textit{Orthant-scan} without sampling is the fastest for the
other four.
We observe that the sampling phase on average scans
only about $5\%$ of the data set, and results in up to 2.55x (1.47x on average) speedup
compared to just running \textit{Orthant-scan}. 
% \shangdi{why on some data sets, orthant-scan is faster than sampling? it is because on those data sets, orthant-scan does not need to scan a lot of points?}
% \yiqiu{Sort of. It's possible that orthant-scan finishes scanning just in one round, and the result is already correct. Some of the data sets may have that property.}
Comparing across different implementations of Welzl's algorithms,
we see that the move-to-front, and the pivoting heuristic implemented
in parallel consistently improve the running times.
Specifically,
\textit{WelzlMtf} is 2.09--13.9x
faster than \textit{Welzl}, and \textit{WelzlMtfPivot} is 3.4--58.6x faster than \textit{Welzl}.
We also see that \textit{Sampling} and \textit{Orthant-scan} are
4.63--34.8x and 2.96--40.3x faster than \textit{WelzlMtfPivot}, respectively.

% We show the scaling vs.\ the input size of our sampling-based algorithm in Appendix~\ref{appendix:hull-seb-vsn}.
% Our implementations can process data sets with up to one billion points. In Figure~\ref{plot:seb-vsn}, we plot the running time of our parallel sampling-based implementation against the number of data points, ranging from one thousand to one billion. We observe a close-to-linear scaling as data size increases.

\input{batch-kdtree/sections/experiments}

%% file: plots/hull2d.tex
\begin{figure*}[t]
\centering
  \includegraphics[width=\textwidth]{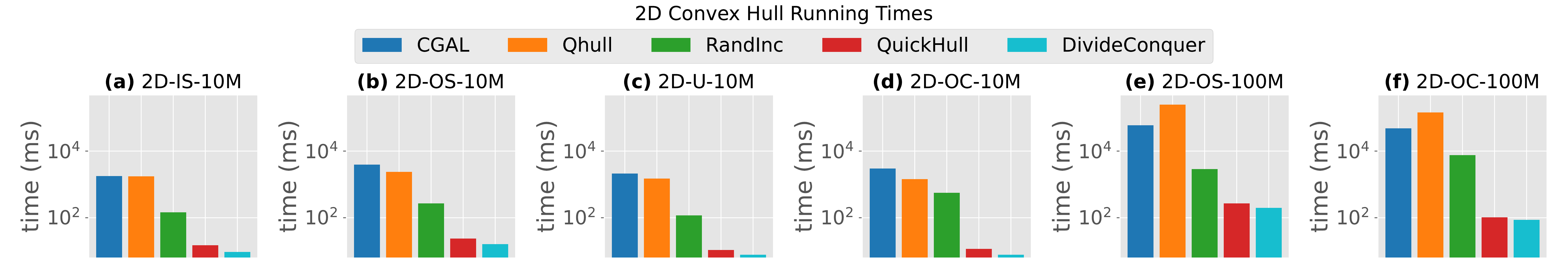}
  \caption{Running times of implementations across different data sets for 2-dimensional convex hull on 36 cores with 2-way hyper-threading.}
  \label{plot:hull2d}
\end{figure*}

%% file: plots/hull3d.tex
\begin{figure*}[t]
\centering
  \includegraphics[width=0.7\textwidth]{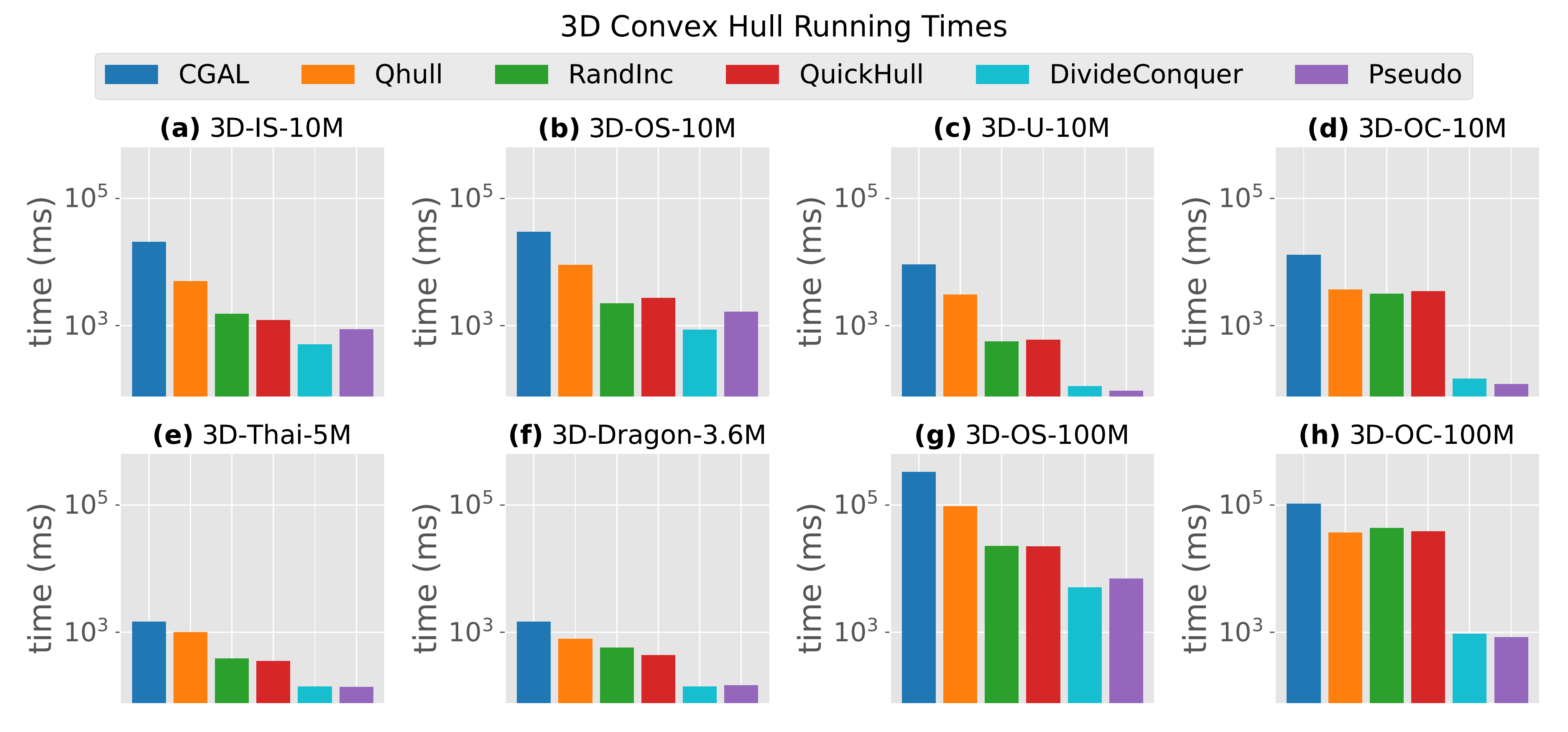}
  \caption{Running times of implementations across different data sets for 3-dimensional convex hull on 36 cores with 2-way hyper-threading.}
  \label{plot:hull3d}
\end{figure*}

%% file: plots/seb.tex
\begin{figure*}[t]
\centering
  \includegraphics[width=\textwidth]{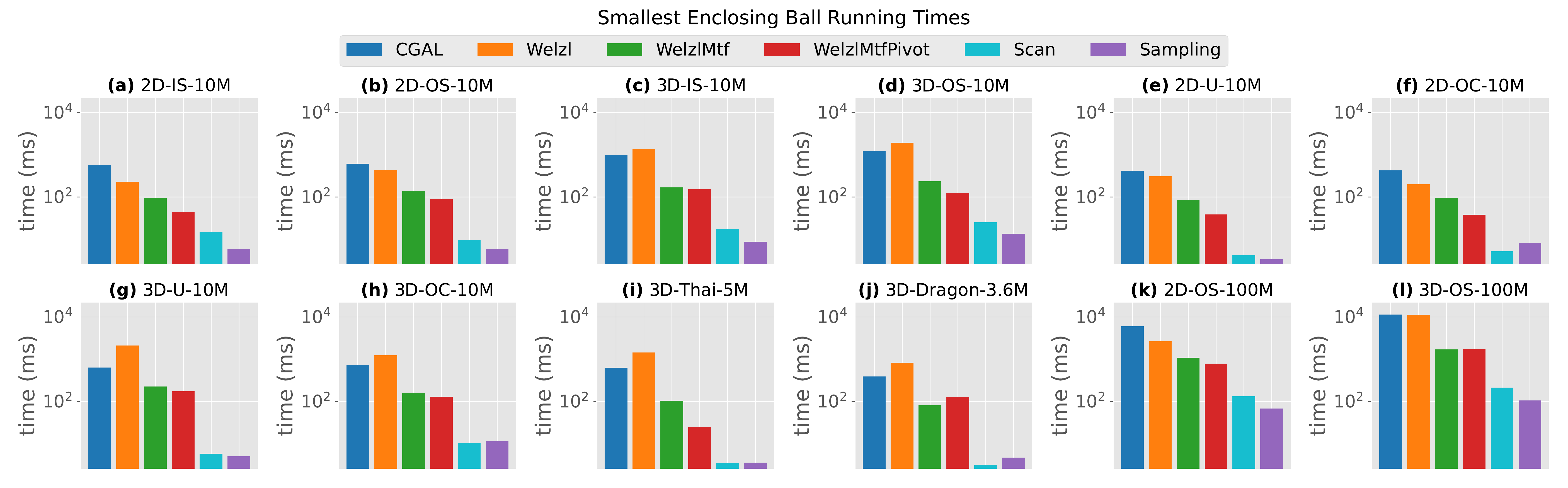}
  \caption{Running times of implementations across different data sets for smallest enclosing ball on 36 cores with 2-way hyper-threading.}
  \label{plot:seb}
\end{figure*}

%% file: batch-kdtree/sections/experiments.tex
\subsection{\ourtree{}}
% \section{Experiments}

We designed a set of experiments to investigate the performance and scalability of \ourtree{} and compare it to two baselines that we also implemented.
\textbf{B1} is a baseline where the \kdtree{} is rebuilt on each batch insertion and deletion in order to maintain balance. This allows for improved query performance (as the tree is always perfectly balanced) at the cost of slowing down updates.
\textbf{B2} is another baseline that inserts points directly into the existing tree structure without recalculating the splits. This results in very fast updates at the cost of potentially skewed trees (which slows down query performance).
\textbf{BDL} is our \ourtree{} described in Section~\ref{sec:ourtree}.
We consider splitting the points based on either using the object median (median among the points along a dimension) or the spatial median (splitting the space along a dimension in half).

\myparagraph{Construction}
% In this benchmark, we measure the time required to construct a tree over each of the data sets. The results using an object median splitting heuristic are shown in Table~\ref{tab:construction-object} and the results using a spatial median splitting heuristic are shown in Table~\ref{tab:construction-spatial}.
Figure~\ref{fig:construction} shows the scalability of the throughput on the 7D-U-10M data set.
As we can see from the results, \textbf{BDL} achieves similar or better performance both serially and in parallel than both \textbf{B1} and \textbf{B2}, and has similar or better scalability than both. With the object median splitting, it achieves up to $34.8\times$ self-relative speedup, with an average self-relative speedup of $28.4\times$. 
We also note that the single-threaded runtimes are faster with spatial median splitting than with object median splitting. This is because spatial median only involves splitting points at each level compared with finding the median for object median, hence it is less expensive to compute; however, we also note that the scalability for spatial median is lower because there is less work to distribute among parallel threads.
The construction of B2 is significantly slower than that of B1, because a separate
memory buffer is allocated at each leaf node in B2 to allow for future insertions.
The construction of the \ourtree{} is faster than both B1 and B2 because splitting the
construction across multiple trees while keeping the number of elements the same
reduces the total work, and provides ample parallelism when running on multiple threads.
% on the other hand, B1 allocates the memory once, and uses parallel splitting to divide
% the data points among leaf nodes.

% \shangdi{why BDL is faster than B1 and B2? and what is B2 in this construction case? Do you mean building the tree by inserting points one by one? and are you measuring the construction time across all batches or just constructing once? }\yiqiu{I will ask Rahul why that's the case. Regarding B2, there's a separate construction function.}\shangdi{how is B2's construciton different from B1?}
% \julian{in theory, the construction performance of B1 and B2 should be exactly the same, and BDL should be slower. we may need to remove this experiment if we cannot figure this out.}
% \yiqiu{Great points. I added the explanation above. There's actually quite a lot of differences in the implementation of the construction functions.} \julian{how about BDL being faster?} \yiqiu{Rahul mentioned higher parallelism due to multiple trees being constructed at the same time, which I think is true due to higher scalability as as shown in the tables we had. I think this does not complete the story as the BDL is also faster on a single thread as shown in the table. I think it actually makes sense because if the total number of points are the same, splitting into multiple trees actually reduce the work (imaging if constructing a tree takes O(n log n), constructing n points into n trees will only take O(n) total). I added some simple explanations above.}
% \input{batch-kdtree/sections/experiments/table-construction}

\input{batch-kdtree/sections/experiments/plot-scalability}

% \subsection{Insertion}\label{exp:insertion}
\myparagraph{Batch Insertion}
In this benchmark, we measure the performance of our batch insertion implementation as compared to the baselines.
% We split this experiment into two separate benchmarks, one to measure the full scalability and performance of our implementation, and one to measure the impact of varying batch sizes. \julian{I am not sure that we have two separate benchmarks.}\yiqiu{Fixed. Forgot to remove this sentence}
%\subsubsection{Scalability}\label{exp:insert-scalability}
We measure the time required to insert 10 batches each containing 10\% of the points in the data set into an initially empty tree for each of our two baselines as well as our \ourtree{}. 

% The results using object median and spatial median splitting heuristics are shown in Tables~\ref{tab:insert-object},~\ref{tab:insert-spatial}, respectively. Figure~\ref{fig:insert-scalability} shows the scalability of the throughput on the 10M points 7D uniform data set.

From Figure~\ref{fig:insert-scalability},
we see that \textbf{B2} achieves the best performance on batched insertions---this is due to the fact that it does not perform any extra work to maintain balance and simply directly inserts points into the existing spatial structure. \textbf{BDL} achieves the second-best performance---this is due to the fact that it does not have to rebuild the entire tree on every insert, but amortizes the rebuilding work across the batches. Finally, \textbf{B1} has the worst performance, as it must fully rebuild on every insertion. Similar to construction, we note that spatial median splitting performs better in the serial case but has lower scalability.  With object median splitting, \textbf{BDL} achieves parallel self-relative speedup of up to $35.5\times$, with an average self-relative speedup of $27.2\times$.

\myparagraph{Batch Deletion}
% In this benchmark, we measure the performance of our deletion implementation as compared to the baselines. Similar to the insertion experiment, we split this experiment into two separate benchmarks, one to measure the scalability and performance of our implementation, and one to measure the impact of varying batch size.
% \subsubsection{Scalability}\label{exp:delete-scalability}
We measure the time required to delete 10 batches each containing 10\% of the points in the data set from an initially full tree for each of our two baselines as well as the \logtree{}.
% The results using an object median splitting heuristic are shown in Table~\ref{tab:delete-object} and the results using a spatial median splitting heuristic are shown in Table~\ref{tab:delete-spatial}. Figure~\ref{fig:delete-scalability} shows the scalability of the throughput on the 10M point 7D uniform data set.
From Figure~\ref{fig:delete-scalability},
we observe that \textbf{B2} has vastly superior performance---it does almost no work other than tombstoning the deleted points so it is extremely efficient. Next, we see that \textbf{BDL} has the second-best performance, as it amortizes the rebuilding across the batches, rather than having to rebuild across the entire point set for every delete. Finally, \textbf{B1} has the worst performance as it rebuilds on every delete. With object median splitting, \textbf{BDL} achieves parallel speedup of up to $33.1\times$, with an average speedup of $28.5\times$.

% \input{batch-kdtree/sections/experiments/table-deletion}

% \subsubsection{Batch Size}\label{exp:delete-batch}
% In this benchmark, we measure the performance of our batch deletion implementation as the size of the batched update varies from 1M points to 5M points. We provide plots of the results for the 2D VisualVar and 7D Uniform datasets in Figures~\ref{fig:batch-delete-2dv},~\ref{fig:batch-delete-7du}, respectively. We note again that \textbf{B2} consistently has the highest throughput, with \textbf{BDL} in second and \textbf{B1} with the lowest throughput. Furthermore, the throughput of \textbf{B1} and \textbf{BDL} increases as the batch size increases; this is true for the same reasons as with insertions. For \textbf{B2}, we observe consistent throughput across batch sizes.

% \subsection{Data-Parallel \knn}
\myparagraph{Data-Parallel \knn{}}
We measure the performance and scalability of our \knn{} implementation as compared to the baselines. % We split this into three separate experiments. 
% \subsubsection{Scalability}\label{exp:knn-scalability}
% In this experiment, we measure the scalability of the \knn{} operation after constructing each data structure over the entire data set (in a single batch).  The results using an object median splitting heuristic are shown in Table~\ref{tab:knn-object} and the results using a spatial median splitting heuristic are shown in Table~\ref{tab:knn-spatial}. Figure~\ref{fig:knn-scalability} shows the scalability of the throughput on the 10M point 7D uniform data set. With the object median heuristic, \textbf{BDL} achieves a parallel speedup of up to $46.1\times$, with an average speedup of $40.0\times$.
As shown in in Figure~\ref{fig:knn-scalability},
the results show that \textbf{B1} and \textbf{B2} have similar performance.
%(\textbf{B2} is slightly faster due to implementation differences \julian{this is confusing because we said earlier that B2 should be slower due to having a more skewed tree. we should explain this more}\yiqiu{B1 is implemented by Rahul, and B2 by me, so there may be some minor differences. My guess is that the \knn{} performances are really the same, the differences are mostly noise. Is it ok to remove the sentence in the parenthesis?}). 
Furthermore, they are both faster than \logtree{}. This is to be expected, because the \knn{} operation is performed directly over the tree after it is constructed over the entire data set in a single batch. Thus, both baselines will consist of fully balanced trees and will be able to perform very efficient \knn{} queries. On the other hand, \textbf{BDL} consists of a set of multiple trees, which adds overhead to the \knn{} operation, as it must be performed separately on each of these individual trees.
% However, as the next two benchmarks show, \textbf{BDL} provides superior performance in the case of a mixed set of dynamic batch inserts and deletes interspersed with \knn{} queries.
% \shangdi{since we don't have enough space to put everything, maybe we should keep the experiments where BDL is superior in the main text?}\yiqiu{We may not have space to include the mixed experiments though. The current experiments may be more representative of the true performance.}
\iffull
In Appendix~\ref{appendix:bdl-varyk}, we show that when the trees are constructed via a set of batch insertions rather than all at once, the performance of \textbf{B2} suffers significantly due to the tree being unbalanced.
\else
In the full version of our paper, we show that when the trees are constructed via a set of batch insertions rather than all at once, the performance of \textbf{B2} suffers significantly due to the tree being unbalanced.
\fi

\myparagraph{Comparison with Zd-tree} We compared with the Zd-tree recently proposed by  Blelloch and Dobson~\cite{dobson2021parallel}. The Zd-tree data structure combines the approach of a \kdtree{} and Morton ordering of the data set, and supports parallel batch-dynamic insertions and deletions, and \knn{}. The implementation currently only supports 2 and 3 dimensional data sets, 
whereas our implementation is not restricted to 2 and 3 dimensions.
We tested their implementation on 3D-U-10M. Using all threads, their implementation takes $0.12$ seconds to construct, and an average of $0.026$ and $0.024$ seconds for insertion and deletion of $10\%$ of the data points, and takes $1.65$ seconds for \knn{}. Our \ourtree{} implementation is $3.3\times$, $23.1\times$, and $45.83\times$ slower, for construction, insertion, and deletion, respectively, but achieves roughly the same speed for \knn{} search. The reason is that the Morton sort used in their implementation is fast and highly optimized for 2 and 3 dimensions; however, extending this technique to higher dimensions would result in overheads due to more bits needed for the Morton ordering.

%% file: batch-kdtree/sections/experiments/plot-scalability.tex
\begin{figure*}
    \begin{subfigure}{0.5\textwidth}
        \centering
        \includegraphics[width=0.85\textwidth]{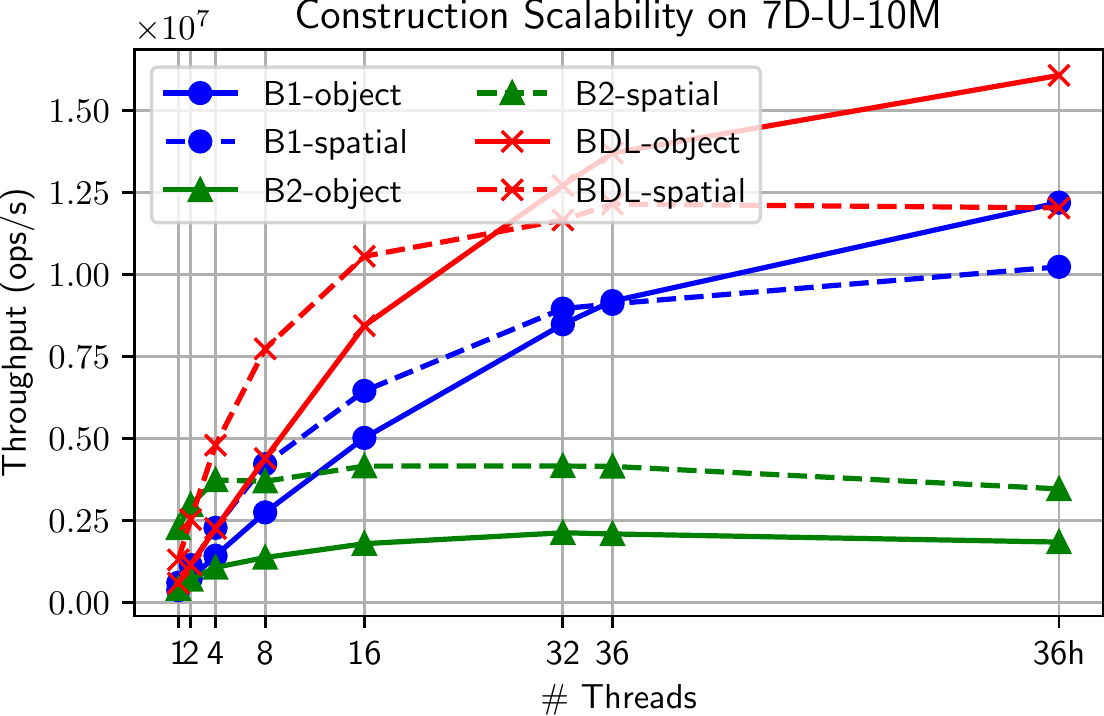}
        \caption{Construction.}
        \label{fig:construction}
    \end{subfigure}
    \begin{subfigure}{0.5\textwidth}
        \centering
        \includegraphics[width=0.85\textwidth]{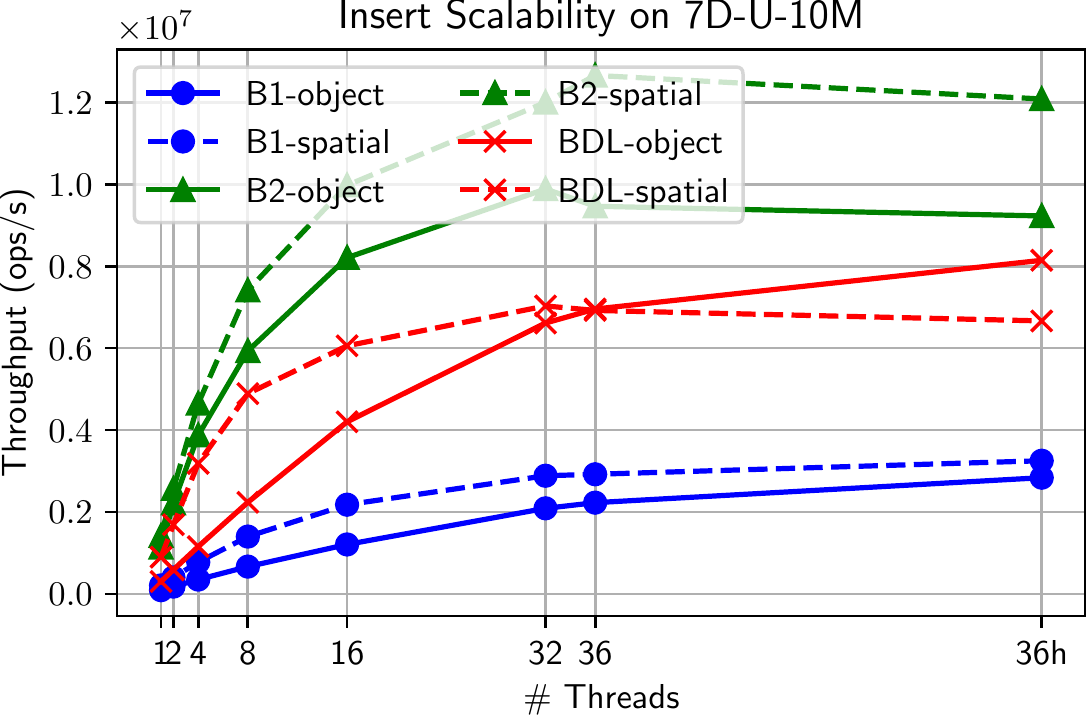}
        \caption{10\% (1M points) Batch Insertion.}
        \label{fig:insert-scalability}
    \end{subfigure}
    \begin{subfigure}{0.5\textwidth}
        \centering
        \includegraphics[width=0.85\textwidth]{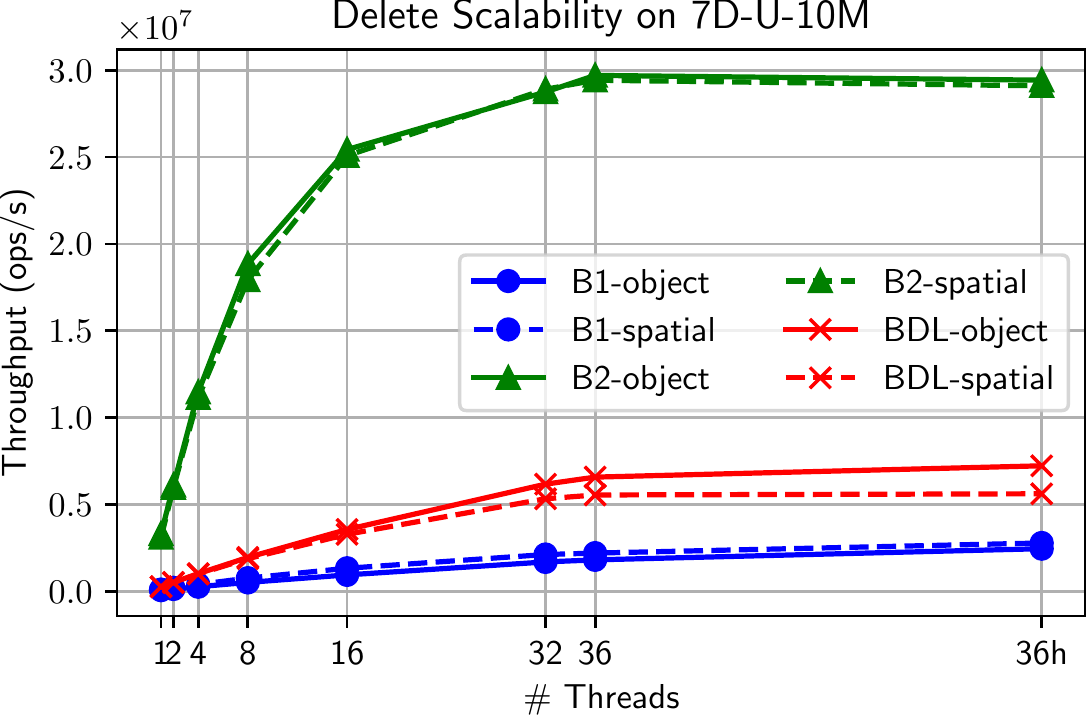}
        \caption{10\% (1M points) Batch Deletion.}
        \label{fig:delete-scalability}
    \end{subfigure}
    \begin{subfigure}{0.5\textwidth}
        \centering
        \includegraphics[width=0.85\textwidth]{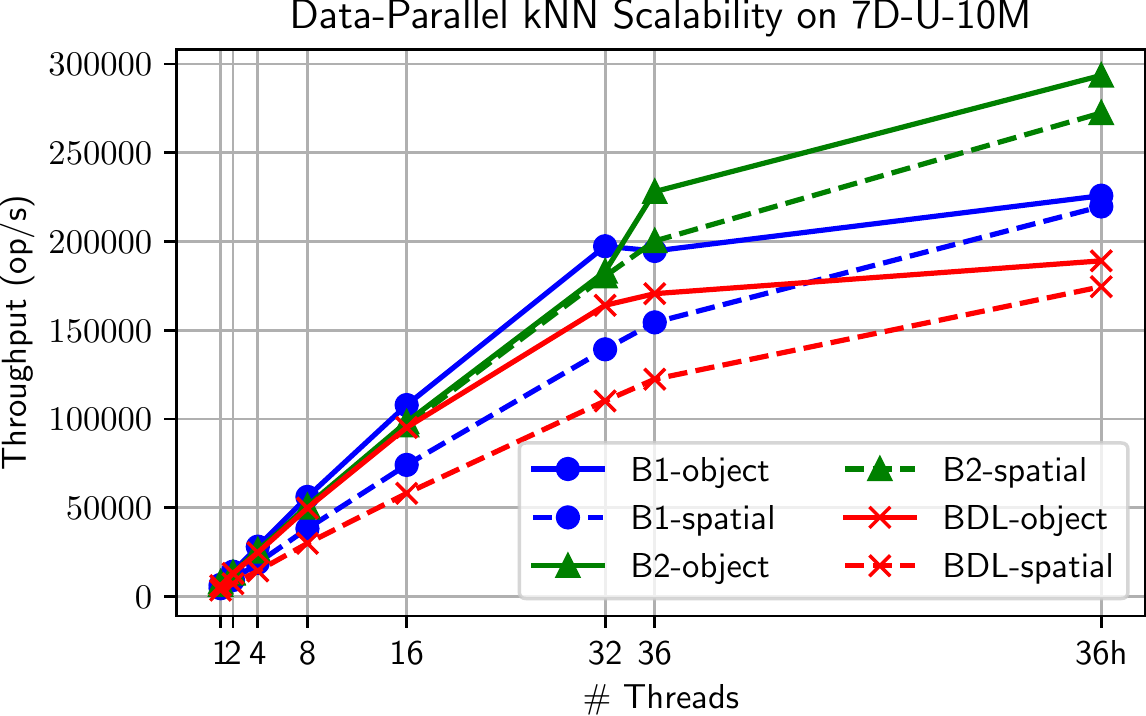}
        \caption{Full (10M points) \knn{} for $k=5$.}
        \label{fig:knn-scalability}
    \end{subfigure}
    \caption{Plot of throughput (operations per second) of batch operations over thread count for both object and spatial median implementations for the 7D-U-10M data set. The prefix of the implementation name refers to the median splitting heuristic. "36h" corresponds to 36 cores with two-way hyper-threading.}
\end{figure*}

%% file: sections/conclusion.tex
\section{Conclusion}

In this paper, we presented ParGeo, a multicore library for computational geometry
containing modules for fundamental tasks including $k$d-tree based spatial search,
spatial graph generation, and algorithms in computational geometry.
We also presented new parallel algorithms, implementations,
and optimizations for convex hull, smallest enclosing ball, and parallel batch-dynamic
$k$d-tree.
We performed a comprehensive experimental study showing that our new implementations achieve significant speedups over prior work and obtain high parallel scalability.

%% file: sections/appendix/hull.tex
\section{Convex Hull} \label{appendix:hull-detailed}
% Content moved to the body of the paper

\hide{
\myparagraph{Overview}
We first give a high-level overview of the algorithm,
whose pseudocode is shown in Figure~\ref{code:hull-high-level}.
Given an ordered set of points $P=\{p_1,p_2,\dots,p_n\}$,
we let $P_r=\{p_1,p_2,\dots,p_r\}$ be the prefix of $P$ of size $r$, and $CH(P_r)$ be the convex hull on $P_r$.
We start the construction by first selecting four
points from $P$ that do not lie on the same plane,
forming a tetrahedron (Line~\ref{line:hull-init}).
We then make these four points the first
four in $P$, and denote the tetrahedron as $CH(P_4)$.
Then, the algorithm proceeds iteratively,
% Then, the algorithm is round-based like the sequential incremental algorithm,
but on each round,
rather than inserting just $p_r$ to form $CH(P_r)$,
we process a batch of points in parallel.
On each round,
let each point outside of $CH(P_{r-1})$ be called a
\defn{visible point}.
We first select a batch of visible points (Line~\ref{line:apex}),
and try to add them to $CH(P_{r-1})$ in parallel in
the same round.

The key challenge of this approach is that some of these
points cannot be processed in parallel due to
concurrent modifications on the shared structures of the
convex polyhedron.
We use a reservation algorithm to resolve these conflicts,
such that we only process the points that modify disjoint
facets of the polyhedron (Lines~\ref{line:reserve-s}--\ref{line:check-e}).
Specifically, each point will perform a \writeMin{} with its ID to reserve all of its visible facets (Lines~\ref{line:reserve-s}--\ref{line:reserve-e}). 
Points that have its id written to all of its visible facets are successful (Lines~\ref{line:check-s}--\ref{line:check-e}).
We then process the successful points in parallel
by enabling them to make concurrent modifications to
$CH(P_{r-1})$ (Line~\ref{line:add-s}--\ref{line:add-e}).
At the end of the round, we run a \pack{}
to filter out points that are no longer visible (Line~\ref{line:pack}).
The algorithm will terminate when there are no more visible points.
}

\myparagraph{Randomized Incremental Algorithm}
Our reservation-based algorithm can be used to implement
the parallel randomized incremental algorithm.
At the start of the algorithm, we randomly permute $P$.
Line~\ref{line:apex} will then take a prefix of the remaining points in $P$ to process.
On each round, we choose a prefix of $c\cdot numProc$ visible
points to perform the reservation, where $c$ is a small constant
and $numProc$ is the number of processors.
Compared with the existing parallelization approach by
Blelloch~\etal~\cite{blelloch2020}, our approach is much simpler
because we avoid the use of complicated data structures.

\myparagraph{Quickhull Algorithm}
Our reservation-based algorithm also applies to the quickhull algorithm.
Specifically, on Line~\ref{line:apex} of the algorithm,
we select a set of visible points that are furthest
from their respective visible facets.
The number of visible points that we choose to
process on each round is again $c\cdot numProc$.

The 3-dimensional quickhull algorithm is one of the most
widely used convex hull algorithms in practice, and so we compare
with some of the existing approaches based on quickhull.
Since concurrent insertions of
visible points creates data races,
existing 3 dimensional implementations
compromise either correctness or scalability.
For Stein~\etal's CudaHull algorithm~\cite{stein2012},
on each round,
the algorithm chooses the furthest point for each facet, and replaces each facet with three
new facets, which is done in parallel on the GPU.
However, such an approach does not produce a convex
polyhedron. Therefore, the implementation
uses the CPU to fix the concave artifacts
produced via ridge rotation at the end of each round.
However, Gao~\etal~\cite{gao2013} pointed out that certain artifacts cannot be fixed by Stein~\etal's algorithm, leaving concavities in the final polyhedron.
Tang~\etal's GPU-based heuristic for convex hull~\cite{tang2012}.
It first generates a ``pseudohull'' polyhedron using a quickhull-like algorithm,
in which the points are removed.
The convex hull is then computed on the remaining points sequentially
on the CPU.
A clear drawback is that the last phase of the algorithm is not parallel, causing a scalability bottleneck for certain data sets.
In comparison, our approach computes a correct convex hull
while also achieving high parallel scalability.
% This is because each round of our algorithm produces a correct convex hull
% on the points processed so far, and do not produce any concavities.
% On the other hand, our method is also end-to-end parallel.
The correctness is because for each visible point, our algorithm considers all the
visible facets and replaces them with new facets.
Meanwhile, Stein~\etal's
algorithm only considers one visible facet for each visible point, and
replaces it directly with three facets, giving rise to concavities during the process.
% \shangdi{in the appendix, we said "we recursively grow each facet into
% three new facet". how is this different from Sten's? Seems that both grow a facet into three new facets}\yiqiu{I actually couldn't find the part you mentioned in the appendix. Where is it? Our reservation based algorithm does not necessarily grow each facets into three}\shangdi{it's in line 708-709 in appendix C. so you mean the difference is that in the last phase where we use the reservation based algorithm, we does not grow each facet into three (bu t in the first phase we do)?}
%We provide more details of the algorithm in Appendix~\ref{appendix:hull-detailed}.

% The quickhull algorithm for $\mathbb{R}^2$ is known
% to be much easier to implement than in $\mathbb{R}^3$.
% In $\mathbb{R}^2$, given an initial convex polygon, each
% facet (edge of the polygon) can be recursively expanded
% by processing the furthest point to replace the facet
% with two new facets. The recursive calls are independent
% to each other, and hence can also be parallelized
% easily~\cite{blelloch1990vector}.

\section{Overhead of Reservation} \label{appendix:hull-overhead}

% \myparagraph{Overhead of Reservations}
Our parallel algorithms are work-efficient since
it does the same amount of asymptotic work as the
sequential counterparts for both the randomized incremental and quickhull algorithms.
%On each round, we choose a batch of $c\cdot numProc$ visible
%points to perform the reservation
%on Line~\ref{line:apex}, where $c$ is a small constant
%and $numProc$ is the number of processors
In addition, given that the expected number of visible
facets associated with each visible point is constant~\cite{deberg2000compgeom},
the amount of extra work done to perform the reservations
on Lines~\ref{line:apex}--\ref{line:check-e} is a constant multiplicative factor
in each round.
%% However, because each visible point can incur the update
%% of more than a constant number of facets in the worst case,
%% the algorithm can have a linear depth.
%% Nonetheless, in practice, the parallelism is sufficiently
%% high given the number of processors available,
%% which we evaluate empirically in the experiments.
When the number of facets is low, instead of using reservations, we process only a single point per round, and we choose the point from
the facet that is visible to the most visible points, which maximizes the volume increase of the convex hull.

% We also apply a optimization to reduce potential contention
% of the reservation.
% When the number of facets in the convex hull is below
% a predetermined constant,
% we only choose one visible point to process per round.
% Specifically, we choose a point from
% the facet that is visible to the most visible points
% The heuristic potentially maximizes the volume
% increase of the convex hull in the step.
% % It also naturally results in abundant work available when redistributing the visible points from old facets to new facets, due to potentially more visible points present.
% In these rounds, we make point redistribution parallel when updating the convex hull on Line~\ref{line:add-e}.

\begin{figure*}[!t]
\centering
\minipage{0.3\textwidth}
  \includegraphics[width=\linewidth]{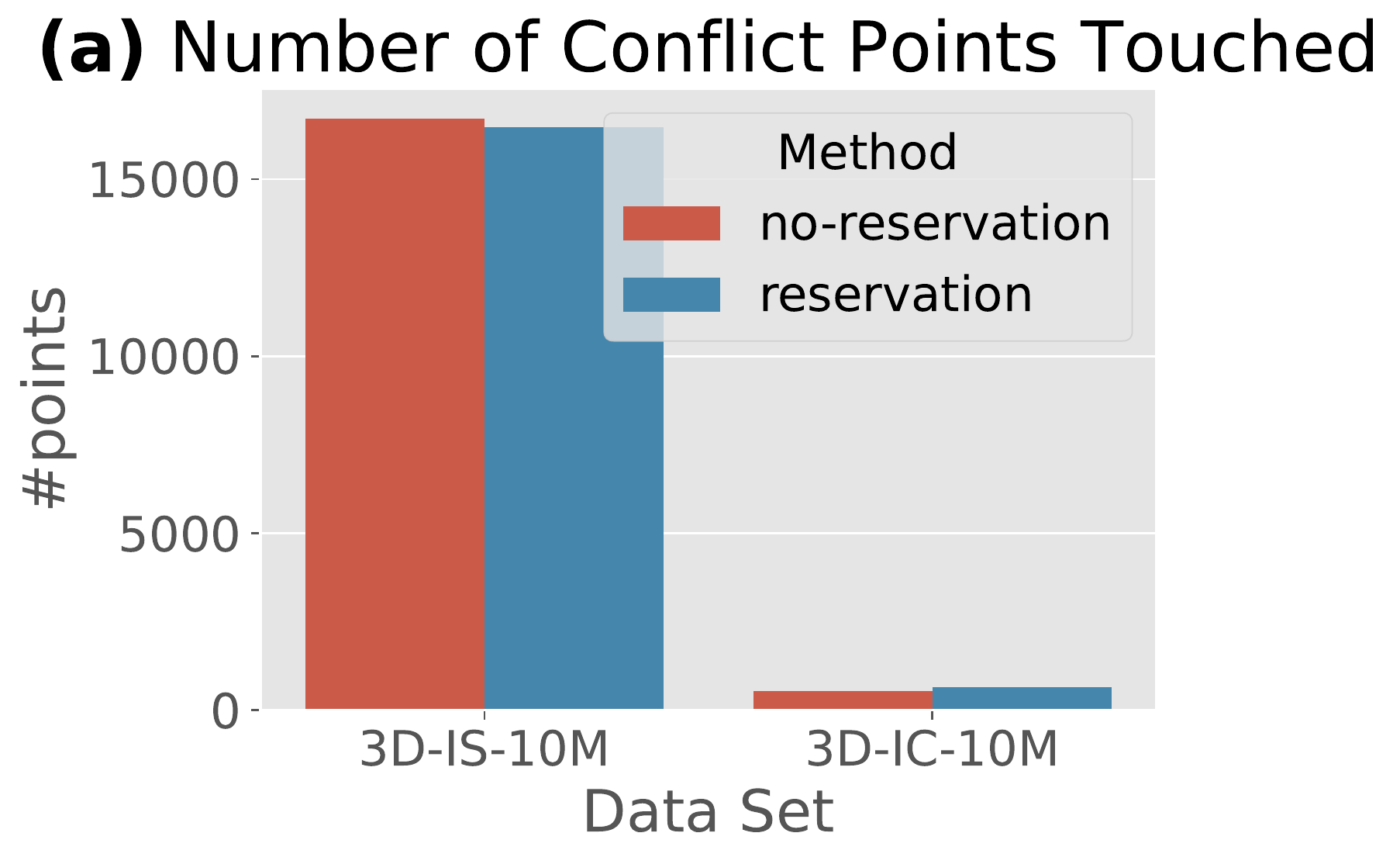}
\endminipage\hfill
\minipage{0.3\textwidth}
  \includegraphics[width=\linewidth]{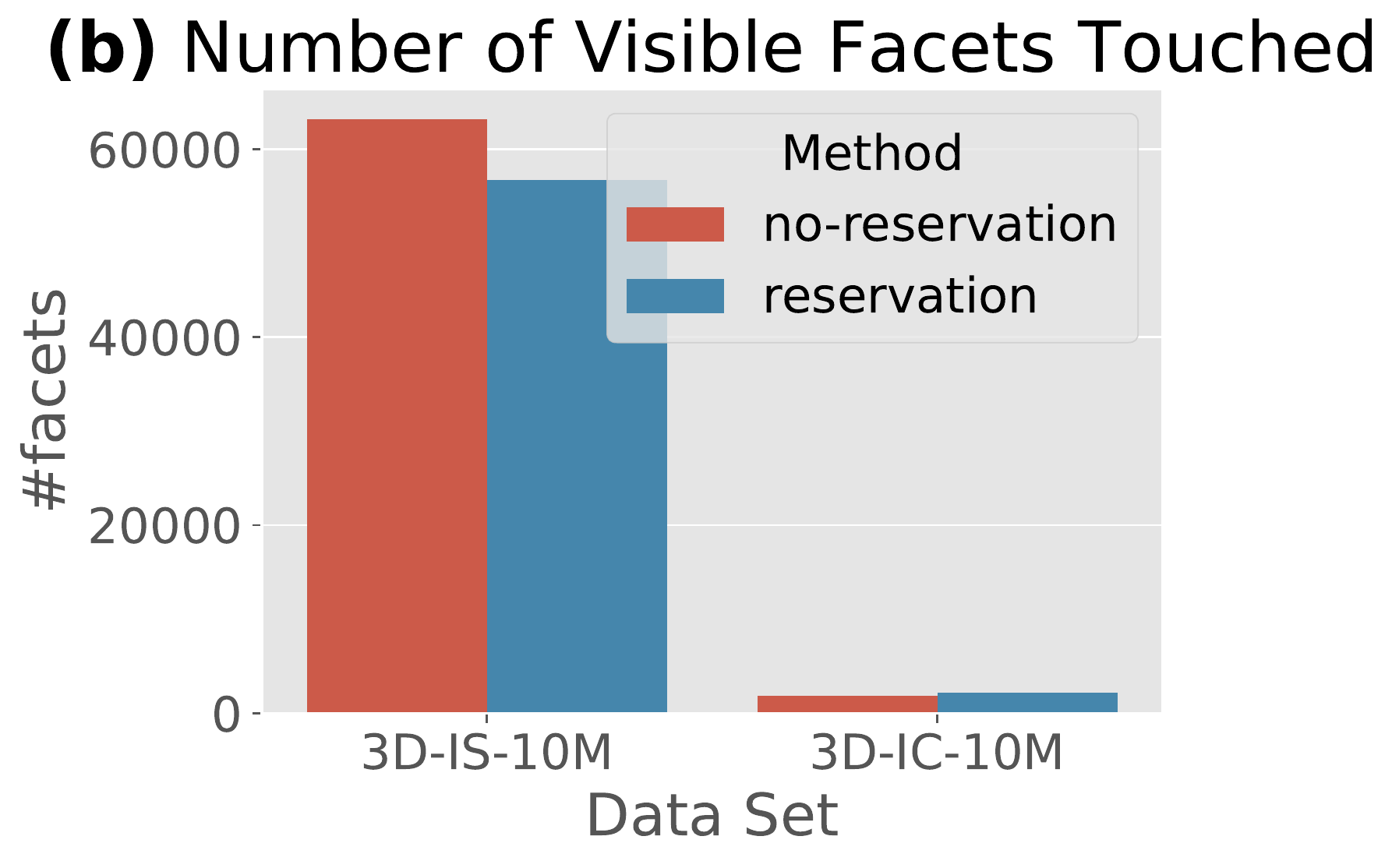}
\endminipage\hfill
\minipage{0.26\textwidth}
  \includegraphics[width=\linewidth]{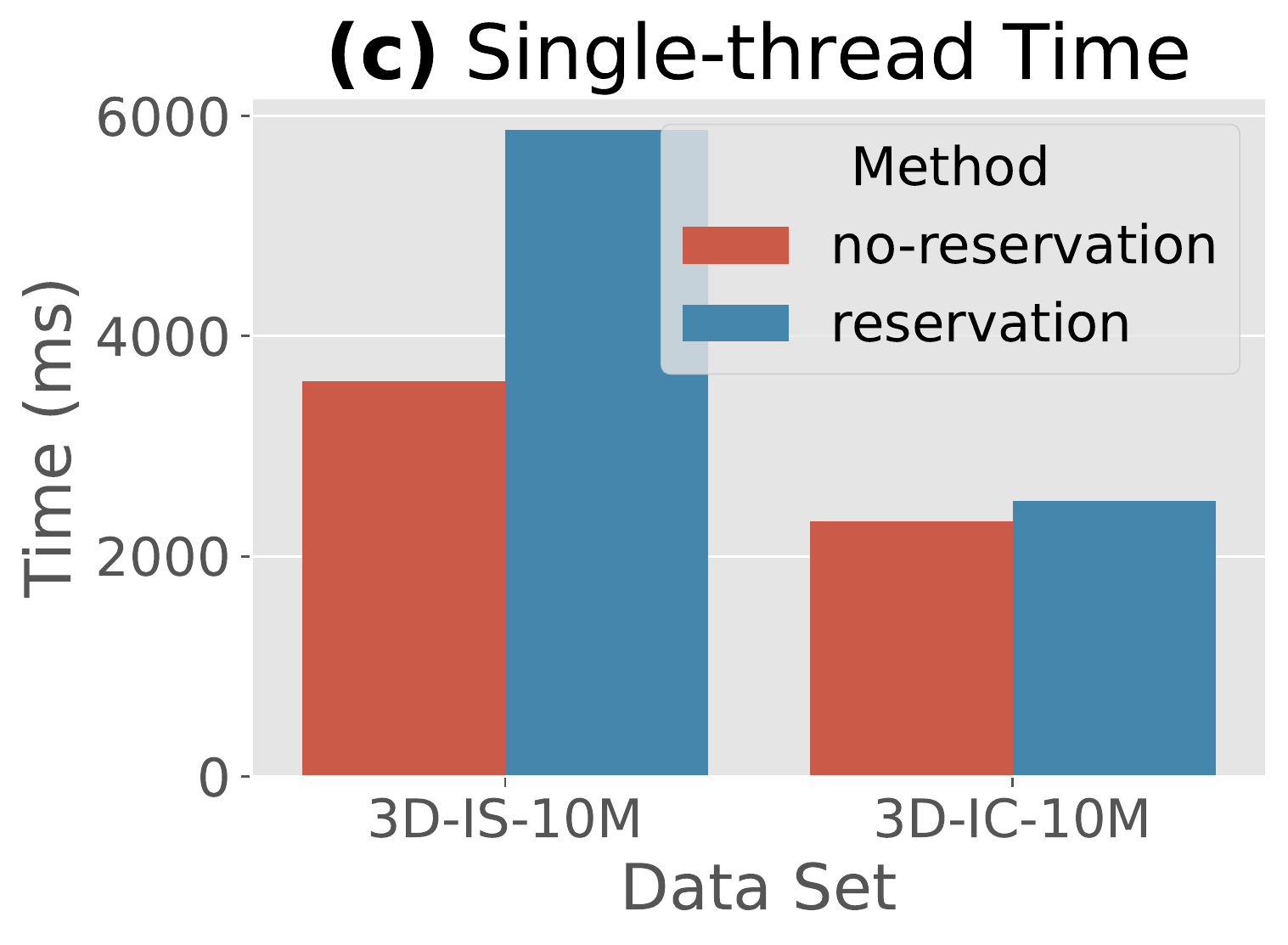}
\endminipage
  \caption{The plots show the overhead of reservation compared with
  without reservation. (a) and (b) show the number of visible
  points and facets touched by the algorithms, respectively. (c) shows the single-thread running time of the algorithms.
  %\julian{label the subfigures (a), (b), and (c) and refer to the letters}\yiqiu{Fixed}
  }\label{fig:hull-overhead}
\end{figure*}

Figure~\ref{fig:hull-overhead} shows an empirical comparison
on the amount of overhead incurred by the reservation algorithm. Specifically, the comparison is between
the reservation-based quickhull algorithm and our optimized sequential quickhull algorithm, both running
on one thread, for data sets containing 10 million points in 3 dimensions (described in Section~\ref{sec:exp}).
The purpose of running on one thread is
to measure the amount of work without parallelism.
The comparison is based on the number of visible points
and facets touched during the algorithm as well as the running time.
As we can see from Figures~\ref{fig:hull-overhead}(a) and (b),
the reservation-based algorithm does not
necessarily cause more points or facets to be touched during
the algorithm as a majority of the reservations
succeed. For example, for the \textit{3D-IS-10M} data set,
the number of visible points and facets touched is
similar to that of the non-reservation algorithm.
For the \textit{3D-IC-10M} data set, the reservation-based approach actually touches fewer
visible points and facets, due to the different order in which the visible points
are selected between the two algorithms.
For both data sets, the reservation algorithm incurs some overhead in
doing the work of reservations, as shown by the single-threaded running times
in Figure~\ref{fig:hull-overhead}(c); however the increase in running time is modest.
This overhead is reasonable since it enables parallelism,
as we will show in our experiments.
% \shangdi{what is $r$ used in the experiment?}\yiqiu{What is r?}\shangdi{I think previously we said the prefix we look at is $r=c*$number of cores. What is the size of the prefix in the experiment?}\yiqiu{The number is suppose to be some constant. But in some version it also changes based on how many points are left. A concrete number for this experiment is missing.}

%% file: sections/appendix/bdl-tree.tex
\section{Detailed \ourtree{}} \label{appendix:bdl-tree}

This section presents more details on operations supported by \ourtree{}. We analyze our algorithms using the work-depth model~\cite{clrs, jaja}. The \emph{\textbf{work}} of an algorithm is the total number of operations used and the \emph{{\textbf{depth}}} is the length of the longest sequential dependence (i.e., the parallel running time).

\input{batch-kdtree/sections/algorithm/single-tree}

\input{batch-kdtree/sections/algorithm/batch-dynamic-insertion}

\input{batch-kdtree/sections/algorithm/batch-dynamic-deletion}

\input{batch-kdtree/sections/algorithm/batch-dynamic-knn}

%% file: batch-kdtree/sections/algorithm/single-tree.tex
\subsection{Static Tree Parallel Algorithms}\label{section:single-alg-top}
\subsubsection{Parallel vEB Construction}
\begin{figure}[!ht]
     \centering
     \includegraphics[width=0.7\textwidth]{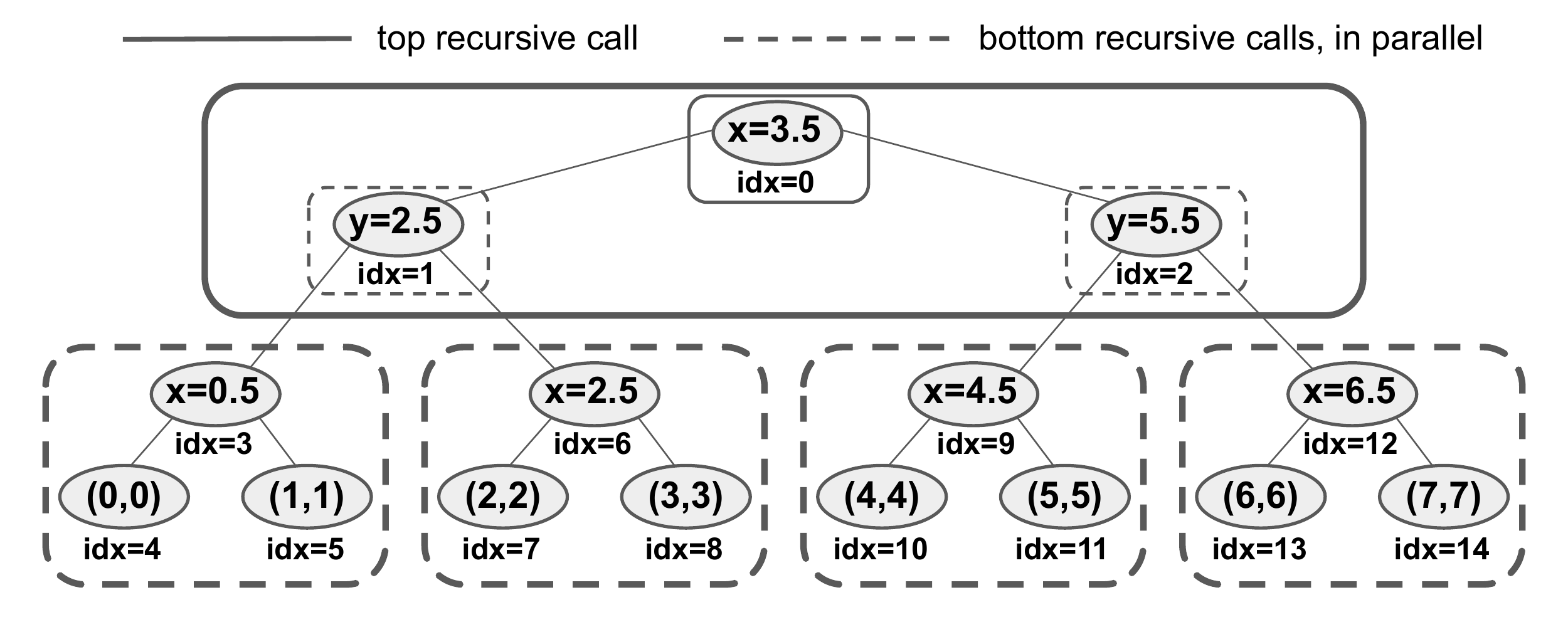}
     \caption{Constructing a vEB \kdtree{} in parallel over 8 2-dimensional points. Note that the top 3 nodes are placed before the remaining 4 bottom subtrees are built in parallel.}
     \label{fig:vebconstruct}
 \end{figure}

\input{batch-kdtree/pseudocode/veb-construct}
The algorithm for parallel construction of the cache-oblivious \kdtree{} is shown in Algorithm~\ref{alg:veb-construct}. The function itself is recursive, and so the top level \textsc{\buildveb{}} function allocates space on line 2 and calls the recursive function \textsc{\buildvebrecursive{}}. Refer to Figure~\ref{fig:vebconstruct} for a graphical representation of this construction.

%\julian{$n$ and $d$ are overloaded. they are used for the number of points and the dimensionality of the data set, respectively, from the prelims. the theorems use $N$ instead of $n$ though. need to make consistent}\rahul{switched $n$, $d$ in the algorithm and made all the theorems use $n$}
The recursive function \textsc{\buildvebrecursive{}} maintains state with 5 parameters: a point set $Q$, a node index $idx$, a splitting dimension $c$, the number of levels to build $l$, and whether it is building the top or bottom of the tree (indicated by $t$). On line 5, we check for the base case---if the number of levels to build is 1, then we have to construct a node. If this is the top of a tree, then this node will be an internal node, so we perform a parallel median partition in dimension $c$ and save it as an internal node. On the other hand, if this is the bottom of the tree, we construct a leaf node that holds all the points in $Q$.
% \yiqiu{Minor point: notation $P$ is slightly confusing as it is overloaded between an entire data set and a subset (see pseudocode)}\rahul{fixed}
%Then, we return. 
Lines 6--9 form the recursive step. In accordance with the exponential layout~\cite{pankaj-co}, we have to first construct the top ``half" of the tree and then the bottom ``half". 
Therefore, on line 6, we compute the number of levels $l_b$ in the bottom portion as the hyperceiling\footnote{The hyperceiling of $n$, denoted as $\lceil\lceil n\rceil\rceil$ is the smallest power of 2 that is greater than or equal to $n$, i.e., $2^{\lceil\log{n}\rceil}$.} of $\frac{l+1}{2}$ and the remaining number of levels $l_t$ in the top portion of the tree as $l - l_b$. On line 7, we recursively build the top half of the tree. 
% \laxman{we should define hyperceiling somewhere---maybe a footnote?}
Then, on line 8, we note that because the top half of the tree is a complete binary tree with $l_t$ levels, it will use $2^{l_t} - 1$ nodes. Therefore, we compute $idx_b = idx + 2^{l_t} - 1$, the node index where the bottom half of the tree should start because the trees are laid out consecutively in memory. Finally, on line 9, we construct each of the $2^{l_t}$ subtrees that fall under the top half of the tree, each with $l_b$ levels. Each of these trees falls into a distinct segment of memory in the array, and so we can perform this construction in parallel across all of the subtrees by precomputing the starting index $idx_i$ for each of the $2^{l_t}$ subtrees.

We trace this process on an example in Figure~\ref{fig:vebconstruct}, in which \textsc{\buildveb{}} is called on a set $P$ of 8 points. This spawns a call to \textsc{\buildvebrecursive{}($P[0:8]$, 0, 0, 4, \textsc{bottom})}. On line 6, we will compute $l_b=2$ and $l_t=2$, and on line 7, we spawn a recursive call to \textsc{\buildvebrecursive{}($P[0:8]$, 0, 0, 2, \textsc{top})}. This call is shown as the solid box around the top 3 nodes in Figure~\ref{fig:vebconstruct}. In this call, we will hit one further level of recursion before laying out the 3 nodes in indices 0, 1, 2. Then, the original recursive call will proceed to line 8, where it will compute $idx_b=3$ as the index to begin laying out the $2^{l_t}=4$ bottom subtrees. Finally, on line 9, we will precompute that the starting indices for the 4 bottom subtrees are $(idx_0, idx_1, idx_2, idx_3) = (3, 6, 9, 12)$. This results in 4 parallel recursive calls, shown in the 4 lower dashed boxes in Figure~\ref{fig:vebconstruct}. Each of these recursive calls internally has one more level of recursion to lay out their 3 nodes.

%\julian{include a running example on a small data set. show one level of recursion, what the top and bottom subproblems are and also what the offsets are}\rahul{done}

\begin{theorem}\label{thm:veb-construction}
The cache-oblivious \kdtree{} with a vEB layout can be constructed over $n$ points in $O(n\log{n})$ work and $O(\log{n}\log{\log{n}})$ depth.
\end{theorem}
\iffull
\begin{proof}
The work bound is obtained by observing that there are $O(\log{n})$ levels in the fully-constructed tree, and the median partition at each level takes $O(n)$ work, giving a total of $O(n\log{n})$ work. 
For the depth bound, at each recursive step we first build an upper tree with size $O(\sqrt{n})$, and then construct the lower trees in parallel, each with size $O(\sqrt{n})$. 
Further, we use an $O(\log{n})$-depth prefix sum to compute $idx_i$ at every level except the base case and an $O(\log{n}\log\log{n})$-depth median partition in the base case. 
Overall, this results in $O(\log{n}\log{\log{n}})$ depth.
\end{proof}
\fi

After the vEB-layout \kdtree{} is constructed, it can be queried as a regular \kdtree{}---the only difference is the physical layout of the nodes in memory. The correctness of this recursive algorithm can be seen through induction on the number of levels. In particular, we form two inductive hypotheses: 
\begin{itemize}[topsep=1pt,itemsep=0pt,parsep=0pt,leftmargin=15pt]
    \item \textsc{\buildvebrecursive{}($Q$, $idx$, $c$, $l$, \textsc{top})} creates a contiguous, fully-balanced binary tree with $l$ levels rooted at memory location $idx$. Furthermore, this binary tree consists of internal \kdtree{} nodes that equally split the point set $Q$ in half at each level. 
    \item \textsc{\buildvebrecursive{}($Q$, $idx$, $c$, $\lceil\log{|Q|}\rceil+1$, \textsc{bottom})} creates a contiguous \kdtree{} with $l$ levels rooted at memory location $idx$.
\end{itemize}
The base cases, with $l=1$, for these inductive hypotheses are explicitly given on line 5. Then, the inductive step follows easily by noting that the definition of hyperceiling implies that the recursive calls on line 9 are all sized such that $l_b = \lceil\log{|Q|_i}\rceil+1$.

\subsubsection{Static Tree Parallel Deletion} %% This is for single tree, insertion is not needed because the static trees are rebuilt in the log tree
\input{batch-kdtree/pseudocode/single-delete}
% \julian{it's strange that deletions are discussed in this section but insertions are not}
% \rahul{in the log structure we don't insert into the single kd-trees, we just rebuild them}
% \laxman{We could also explain before going into Section III.A that we will only talk about construction and deletion for the single-tree algorithms}
The algorithm for parallel deletion from a single \kdtree{} is shown in Algorithm~\ref{alg:single-delete}. The function itself is recursive, so the top level \textsc{\eraseS{}} calls the subroutine \textsc{\eraseSrecursive{}} on the root node on line 2.

The recursive function \textsc{\eraseSrecursive{}} acts on one node at a time, represented by the index $idx$. On line 4, it checks for the base case---if the current node is a leaf node, it simply performs a linear scan to mark any points in the leaf node that are also in $Q$ as deleted. 
% \julian{$P$ is overloaded. we used it for the entire point set. let's call it something else, like $D$?}\rahul{fixed}
Then, it returns \textsc{NULL} if the entire leaf was emptied; otherwise, it returns the current node $idx$. Lines 5--7 represent the recursive case. First, on line 5, we perform a parallel partition of $Q$ around the current node's splitting hyperplane. We refer to the lower partition as $Q_{l}$ and the upper partition as $Q_{r}$. On line 6, we recurse on the left and right subtrees in parallel, passing $Q_{l}$ to the left subtree and $Q_{r}$ to the right. Finally, line 7 updates the tree structure. We always ensure that every node has 2 children in order to flatten any unnecessary tree traversal. The return value of \textsc{\eraseSrecursive{}} indicates the node that should take the place of $idx$ in the tree (potentially the same node)---a return value of \textsc{NULL} indicates that the entire subtree rooted at $idx$ was removed. So, if both the left and right child are removed, then we can remove the current node as well by returning \textsc{NULL}. On the other hand, if neither the left or right child are removed, then the subtree is still intact, and we simply reset the left and right child pointers of the current node and return the current node $idx$, indicating that it was not removed. Finally, if exactly one of the children was removed, then we remove the current node as well and let the remaining child connect directly to its grandparent---in this way, we remove an unnecessary internal splitting node. We do this by simply returning the non-\textsc{NULL} child, signaling that it will take the place of the current node in the \kdtree{}. 
%\julian{add explanation for Line 7. also show on running example.}\rahul{need to do running example}

\begin{theorem}
Deleting a batch of $B$ points from a single \kdtree{} constructed over $n$ points can be done in $O(B\log{n})$ work and $O(\log{B}\log{n})$ depth in the worst case.
\end{theorem}

\iffull
\begin{proof}
We can see the work bound by noting that each of the $B$ points traverse down $O(\log{n})$ levels as part of the algorithm. For the depth, note that in the worst-case the parallel partition at each level operates over $O(B)$ points at each level. Because parallel partition has logarithmic depth, this would result in a worst-case $O(\log{B})$ depth at each of the $O(\log{n})$ levels, giving the overall depth of $O(\log{B}\log{n})$. 
% \julian{partition should be $O(\log B\log\log B)$ depth?}\rahul{this is just partitioning around a value, not around the median - i thought we can do this in $O(\log n)$ depth with prefix sum?}
\end{proof}
\fi

\subsubsection{Data-Parallel \knn{}}
% \input{pseudocode/single-knn}
% \todo{How detailed should this be? its kind of already existing work}\yiqiu{I saw that you described it in the implementation section, so maybe we don't have to talk about it here.}\rahul{should we keep the pseudocode?}\yiqiu{The sequential knn pseudocode seems to be a repetition of the description in the prelim. Maybe you could turn the pseudocode into a description and merge it with previous descriptions. (you could keep the parallel algorthm for knn described later though)}

%We omit the pseudocode for \knn{} searches on \kdtrees{} as it is well-studied~\cite{bentley1975}. 

We execute our \knn{} searches in a data-parallel fashion by parallelizing across all of the query points in a batch. The \knn{} search for each point is executed serially.
% We will give a slightly more detailed description of the approach introduced in Section~\ref{prelim-knn}.
We implement a ``\knn{} buffer", a data structure that maintains a list of the current $k$-nearest neighbors and provide quick insert functionality to test and insert new points if they are closer than the existing set. The data structure maintains an internal buffer of size $2k$. To insert a point, it simply adds that point to the end of the buffer. If the buffer is filled up, then it uses a serial selection algorithm to partition the buffer around the $k$-th nearest element and clears out the remaining $k$ elements. This achieves a serial amortized $O(1)$ runtime (because the selection partition step is $O(k)$ and is only performed for every $k$ insertions).

To implement batched \knn{} on the \kdtree{}, we perform a \knn{} search for each individual point in parallel across all the points. We now describe the \knn{} method (\textsc{\knnserial{}}) for a single point $p$. We first allocate a \knn{} buffer for the point. Then, we recursively descend through the \kdtree{} searching for the leaf that $p$ falls into. When we find this leaf, we add all of the points in the leaf to the \knn{} buffer. Then, as the recursion unfolds, we check whether the \knn{} buffer has $k$ points. If it does not, we add all the points in the sibling of the current node to the \knn{} buffer to try to fill up the buffer with nearby points as quickly as possible to improve our estimate of the $k$-th nearest neighbor. 
%\julian{it's not clear to me why we add points in the sibling to the buffer. we are going to search the sibling later if it is in our bounding box, so we should be able to add those points then.}\rahul{as we unwind, we try to fill up $k$ points as quickly as possible so that we can construct an initial bounding box - in practice this case should not be used many times - only until the initial $k$ neighbors have been added to the buffer}
Otherwise, we use the current distance of the $k$-th nearest neighbor to prune subtrees in the tree. In particular, if the bounding box of the current subtree is entirely contained within the distance of the $k$-th nearest neighbor, we add all points in the subtree to the \knn{} buffer. If the bounding box is entirely disjoint, then we prune the subtree. Finally, if they intersect, we recurse on the subtree.

\iffull
\begin{theorem}
 For a constant $k$, \knn{} queries over a batch of $B$ points can be performed over a single \kdtree{} containing $n$ points in worst-case $O(Bn)$ work and worst-case $O(n)$ depth.
\end{theorem}
\begin{proof}
In the worst-case, we have to search the entire tree, of size $O(n)$, resulting in total work of $O(Bn)$ (due to the amortized $O(1)$ insert cost for \knn{} buffers) and depth of $O(n)$, as the queries are done in parallel over the batch, but each search is serial. 
\end{proof}
\fi

As noted by Bentley~\cite{bentley1975} and Friedman \etal{}~\cite{friedman1977-kdtree-nn}, the work for a single nearest-neighbor query on a \kdtree{} is empirically found to be logarithmic in $n$, so the experimental runtime and scalability are much better than suggested by the worst-case bounds.
% \subsubsection{Dual-Tree \knn{}}
% The pseudocode for this algorithm is given in \cite{esmt}. We parallelize this algorithm and test it experimentally.\todo{How much should we say here? I think we should just mention that we tried it and give some of the experimental results to establish that it did not work well.}

%% file: batch-kdtree/pseudocode/veb-construct.tex
\begin{algorithm}[!t]
\caption{Parallel vEB-layout \kdtree{} Construction}\label{alg:veb-construct}
\hspace*{\algorithmicindent} \textbf{Input}: Point Set $P$ \\
\hspace*{\algorithmicindent} \textbf{Output}: \kdtree{} over $P$, laid out with the vEB layout on a contiguous memory array of size $2|P| - 1$.
\begin{algorithmic}[1]
\Procedure{\buildveb{}}{$P$}
    \State Allocate $2|P| - 1$ nodes in contiguous memory. The tree nodes will be laid out in this space.
    \State \textsc{\buildvebrecursive{}($P$, 0, 0, $\lfloor\log(|P|)\rfloor$+1, bottom)}
\EndProcedure
\Procedure{\buildvebrecursive{}}{$Q$, $idx$, $c$, $l$, $t$}
\Statex $idx$: current node index in the memory array
\Statex $c$: current dimension to split on
\Statex $l$: number of levels to build
\Statex $t$: whether we are building the top or bottom of a tree
\State If we hit the base case $n=1$, then we construct a node at $idx$. If $t$ is \textsc{top}, then we perform a parallel median partition on $Q$ in dimension $c$ and record this split as an internal node. Otherwise, we create a leaf node that represents the points in $Q$.
\State Compute $l_b = \left\lceil\left\lceil \frac{l+1}{2} \right\rceil\right\rceil$ and $l_t = l - l_b$ (vEB layout). 
\State Recursively build the top half of the tree with \textsc{\buildvebrecursive{}($Q$, $idx$, $c$, $l_t$, top)}.
\State Compute $idx_b = idx + 2^{l_t}-1$ as the offset where the top half of the tree was just laid out.
\State Construct the $2^{l_t}$ lower subtrees in parallel with \textsc{\buildvebrecursive{}($Q_i$, $idx_i$, $(c+n_t)\mod{d}$, $l_b$, $t$)} where $Q_i$ is the subarray of points that are held by the parent of this subtree and $idx_i$ is the index at which this subtree is to be placed (precomputed with a parallel prefix sum).% \shangdi{define $n_t$ and $d$.}
\EndProcedure
\end{algorithmic}
\end{algorithm}

%% file: batch-kdtree/pseudocode/single-delete.tex
\begin{algorithm}[!t]
\caption{Parallel $kd$-Tree Deletion}\label{alg:single-delete}
\hspace*{\algorithmicindent} \textbf{Input}: Point Set $P$
% \hspace*{\algorithmicindent} \textbf{Output}: $kd$-Tree over $P$, laid out with a binary-heap layout on a preallocated contiguous memory space of size $2P.len - 1$.
\begin{algorithmic}[1]
\Procedure{\eraseS{}}{$P$}
    \State \textsc{\eraseSrecursive{}($P$, 0)}
\EndProcedure
\Procedure{\eraseSrecursive{}}{$Q$, $idx$}
\Statex $idx$: current node index
\State If the current node is a leaf node, mark any points in the leaf node that are also in $Q$ as deleted. If all of the points in the current leaf are deleted, return \textsc{NULL}. Otherwise, return the current $idx$.
\State Otherwise, perform a parallel partition on $Q$ around the split represented by the current node. Let $Q_{l}, Q_{r}$ be the resulting left and right arrays, respectively, after the partition.
\State Then, recurse on the children in parallel with \textsc{\eraseSrecursive{}($Q_{l}$, $idx_{l}$)}, \textsc{\eraseSrecursive{}($Q_{r}$, $idx_{r}$)}, where $idx_{l}$ and $idx_{r}$ are the IDs of the left and right children, respectively.
\State If neither of the recursive calls return \textsc{NULL}, reset the left and right children to be the results of these calls and return the current node. If both of the recursive calls return \textsc{NULL}, return \textsc{NULL}. 
% (this effectively deletes the current node and its entire subtree).
If one of the recursive calls returns \textsc{NULL} and the other does not, return the non-\textsc{NULL} node.
% (effectively removing the current node but not its subtree).
\EndProcedure
\end{algorithmic}
\end{algorithm}

%% file: batch-kdtree/sections/algorithm/batch-dynamic-insertion.tex
% \subsection{Batch-Dynamic Parallel Algorithms}\label{section:log-alg-top}
% This section describes our algorithms for supporting batch-dynamic updates on \ourtree{}s.

% \yiqiu{Pseudocode is too low level, and contains dangling references, fix, make it self-contained}
% \yiqiu{Also try to move out the example and be more high level in the descriptions}
% \yiqiu{can use myparagraph instead of subsections}

\subsection{Parallel Insertion} \label{appendix:bdl-insert}

\input{batch-kdtree/pseudocode/log-insert}

Insertions are performed in the style of the logarithmic method~\cite{bentley-logarithmic-1, bentley-logarithmic-2}, with the goal of maintaining the minimum number of full trees within \logtree{}. Thus, upon inserting a batch $B$ of points, we rebuild larger trees if it is possible using the existing points and the newly inserted batch.
This is implemented as shown in Algorithm~\ref{alg:log-insert}, and depicted in Figure~\ref{fig:bdl-insert}.
% This is implemented as depicted in Figure~\ref{fig:bdl-insert}.

First, on line 2, we build a bitmask $F$ of the current set of full static trees in the logarithmic structure. Then, on line 3, because the buffer \kdtree{} has size $X$, we can add $|P|/X$ to $F$ to compute a new bitmask $F_{new}$ of full trees that would result if we added $|P|$ points to the tree structure. As an implementation detail, note that we first add $|P| \mod{X}$ points to the buffer \kdtree{}---if we fill up the buffer \kdtree{}, then we gather the $X$ points from it and treat them as part of $P$, effectively increasing the size of $P$ by $X$. Then, on line 4, taking the bitwise difference between these two bitmasks gives the set of trees that should be consolidated into new larger trees---specifically, any tree that is set in $F_{new}$ but not in $F$ must be constructed from trees that are set in $F$ but not in $F_{new}$. After determining which trees should be combined into new trees, on line 5 we construct all the new trees in parallel---in parallel for each new tree to be constructed, we deconstruct and gather all the points from trees that are being combined into it and then we construct the new tree over these points and any additional required points from $P$ using Algorithm~\ref{alg:veb-construct}.

Refer to Figure~\ref{fig:bdl-insert} for an example of this insertion method (suppose for this example that $X>2$). In Figure~\ref{fig:bdl-insert-0}, the \ourtree{} contains $X$ points, giving a bitmask of $F=1$ (because only the smallest tree is in use). If we insert $X+1$ points, then we put one node in the buffer tree and compute $F_{new} = 1 + \frac{X}{X} = 2$, and so we have to deconstruct static tree 0 and build static tree 1, as shown in Figure~\ref{fig:bdl-insert-1}. Then, if we insert $X+1$ points again, then we again put one point in the buffer tree and compute $F_{new}=2 + \frac{X}{X} = 3$, and so we simply construct tree 0 on the $X$ new points (leaving tree 1 intact), as seen in Figure~\ref{fig:bdl-insert-2}. Finally, if we then insert $X-1$ points, we note that this would fill the buffer up, so we take 1 point from the buffer and insert $X$ points; then, $F_{new} = 3 + \frac{X}{X} = 4$, and so we deconstruct trees 0, 1 and construct tree 2, as seen in Figure~\ref{fig:bdl-insert-3}.

% \input{batch-kdtree/sections/algorithm/fig-insert-example}

%% Deletion moved to Appendix
We refer the readers to Appendix~\ref{sec:bdl-parallel-delete} for the parallel algorithm
for batch deletion.

%% file: batch-kdtree/pseudocode/log-insert.tex
\begin{algorithm}[!t]
\caption{Parallel \ourtree{} Batch Insertion}\label{alg:log-insert}
% 
%\hspace*{\algorithmicindent} \textbf{Input}: Point Set $P$ 
% \hspace*{\algorithmicindent} \textbf{Output}: $kd$-Tree over $P$, laid out with a binary-heap layout on a preallocated contiguous memory space of size $2P.len - 1$.
\begin{algorithmic}[1]
\Procedure{\insertlog{}}{$P$}
\State Build an integer bitmask $F$ that represents the static trees within the logarithmic tree structure that are currently filled using 1's, and the trees that are empty using 0's.
\State Compute $F_{new} = F + \frac{|P|}{X}$, where $X$ is the buffer tree size. This is the new bitmask of trees that should be filled.
\State Based on the difference between $F$ and $F_{new}$, determine which trees should be combined into larger trees.
\State Gather the relevant points and construct all the new trees in parallel using \textsc{\buildveb{}} (or \textsc{\buildbhl{}} for the buffer tree).
\EndProcedure
\end{algorithmic}
\end{algorithm}

%% file: batch-kdtree/sections/algorithm/batch-dynamic-deletion.tex
\subsection{\ourtree{} Parallel Deletion} \label{sec:bdl-parallel-delete}
\input{batch-kdtree/pseudocode/log-delete}
When deleting a batch of points, the goal is to maintain balance within the subtrees. Thus, if any subtree decreases to less than half of its full capacity, we move all the points down to a smaller subtree in order to maintain balance. As seen in Algorithm~\ref{alg:log-delete}, this is implemented as a three-step process.

On line 2, we call a parallel bulk erase subroutine on each of the individual trees in parallel in order to actually erase the points from the trees. On line 3, we scan the trees in parallel and collect the points from all trees which have been depleted to less than half of their original capacity. Finally, on line 4, we use the \textsc{\insertlog{}} routine to reinsert these points into the structure.

We provide the bounds for parallel insertion and deletion below.
%and refer the reader to Yesantharao \etal~\cite{yesantharao2021parallel} for the analysis.

\begin{theorem}
Given an \ourtree{} that was created using only batch insertions and deletions, each batch of $B$ updates takes
$O(B\log^2(n+B))$ amortized work and $O(\log{(n+B)}\log\log{(n+B)})$ depth, where $n$ is the number of points in the tree before applying the updates.
\end{theorem}

% \iffull
\begin{proof}
We first argue the work for only performing insertions starting from an empty \ourtree{}. In the worst case, points are added to the structure one by one. Then, as similar to the analysis by Bentley~\cite{bentley-logarithmic-1}, the total work incurred is given by noting that the number of times the $i$'th tree is rebuilt when inserting $m$ points one by one is $O(2^{\log{m}-i})$. Then, summing the total work gives $O(\sum_{i=0}^{\log{m}}{2^{i}i2^{\log{m}-i}}) = O(m\log^2{m})$, where we use the work bound from Theorem~\ref{thm:veb-construction}. After inserting a batch of size $B$, we have $n+B$ points in the \ourtree{}, and so the amortized work for the batch is $O(B\log^2 (n+B))$. Now, if deletions occurred prior to a batch insertion, and the current \ourtree{} has $n$ points, there still must have been $n$ previous insertions (since we started with an empty data structure), and so the work of this batch can still be amortized against those $n$ insertions. We now argue the depth bound. When a batch inserted into the tree, the points from smaller trees can be gathered in worst-case $O(\log{(n+B)})$ depth (if all the points must be rebuilt) and the rebuilding process takes worst case $O(\log{(n+B)}\log\log{(n+B)})$ depth, using the result from Theorem~\ref{thm:veb-construction}.

The initial step of deleting the batch of points from each of the underlying \kdtrees{} incurs $O(B\log^2{n})$ work (there are $O(\log{n})$ \kdtrees{}, each taking work $O(B\log{n})$) and depth $O(\log{B}\log{n})$. Then, collecting the points that need to be reinserted can be done in worst-case depth $O(\log{(n+B)})$ and the reinsertion takes $O(\log{(n+B)}\log\log{(n+B)})$ depth, from before. Overall, the depth is $O(\log{(n+B)}\log\log{(n+B)})$. The amortized work for reinserting points in trees that are less than half full is $O(B\log^2{n})$, as every point we reinsert can be charged to a deletion of another point, either from this batch or from a previous batch. This is because for a tree that is half full, there must be at least as many deletions from the tree as the number of points remaining in the tree.
\end{proof}
% \fi

%% file: batch-kdtree/pseudocode/log-delete.tex
\begin{algorithm}[!t]
\caption{Parallel \ourtree{} Batch Deletion}\label{alg:log-delete}
\hspace*{\algorithmicindent} \textbf{Input}: Point Set $P$
% \hspace*{\algorithmicindent} \textbf{Output}: $kd$-Tree over $P$, laid out with a binary-heap layout on a preallocated contiguous memory space of size $2P.len - 1$.
\begin{algorithmic}[1]
\Procedure{\eraselog{}}{$P$}
\State In parallel, delete $P$ from each of the underlying trees which is nonempty by calling \textsc{\eraseS{}($P$)} on each of these trees.
\State In parallel, gather the points from any trees that drop to below half of their original capacity into a set $R$.
\State Call \textsc{\insertlog{}($R$)} to reinsert these points into the log-tree structure.
\EndProcedure
\end{algorithmic}
\end{algorithm}

%% file: batch-kdtree/sections/algorithm/batch-dynamic-knn.tex
\subsection{Data-Parallel $k$-NN}\label{alg:dpknn}

% \input{batch-kdtree/pseudocode/log-knn}
% In the data-parallel \knn{} implementation, we parallelize over the set of points given to search for nearest neighbors. First, on line 2, we allocate a \knn{} buffer for each of the points in $S$. Then, for each of the non-empty trees in \logtree{}, we call the data-parallel \knn{} subroutine on the individual tree, passing in the set $S$ of points and the set of \knn{} buffers. Because we reuse the same set of \knn{} buffers for each underlying \knn{} call, we eventually end up with the $k$-nearest neighbors across all of the individual trees for each point in $S$.
In the data-parallel \knn{} implementation, we parallelize over the set of points given to search for nearest neighbors. First, we allocate a \knn{} buffer for each of the points in $S$. Then, for each of the non-empty trees in \logtree{}, we call the data-parallel \knn{} subroutine on the individual tree, passing in the set $S$ of points and the set of \knn{} buffers. Because we reuse the same set of \knn{} buffers for each underlying \knn{} call, we eventually end up with the $k$-nearest neighbors across all of the individual trees for each point in $S$.

% \iffull
% \begin{theorem}
% For a constant $k$, \knn{} queries over a batch of $B$ points over the $n$ points in the \ourtree{} can be performed in $O(Bn)$ work and $O(n)$ depth.
% \end{theorem}
% \begin{proof}
% These bounds follow directly from the bounds of the underlying individual \knn{} calls. The \knn{} routine on the $i$'th underlying tree, with size $n_i$, has worst-case work $O(Bn_i)$ and depth $O(n_i)$. Summing over all $i$ gives the bounds.  
% \end{proof}
% \fi

%% file: sections/appendix/experiments.tex
% \section{Other Experiments}

% \subsection{Scaling Convex Hull and Smallest Enclosing Ball with Input Size}\label{appendix:hull-seb-vsn}

% \input{plots/hull-vsn}

% Our convex hull implementations can process data sets with up to one billion points, which is 10x larger than the largest data set used in testing existing implementations for parallel architectures~\cite{stein2012,tang2012,gao2013,srikanth2009parallelizing,masnadi2020}. In Figure~\ref{plot:hull-vsn}, we plot the running time of our parallel quickhull implementation against the number of data points, ranging from one thousand to one billion. We observe a close-to-linear scaling as the data size increases.

% \input{plots/seb-vsn}

% Our smallest enclosing ball implementations can process data sets with up to one billion points. In Figure~\ref{plot:seb-vsn}, we plot the running time of our parallel sampling-based implementation against the number of data points, ranging from one thousand to one billion. We observe a close-to-linear scaling as data size increases.

\section{\knn{} Performance on Incrementally Constructed Trees}\label{appendix:bdl-varyk}

\input{batch-kdtree/sections/experiments/plot-knn-varyk}

In this experiment, we benchmark the throughput of the \knn{} operation on 36 cores with hyper-threading as $k$ varies from 2 to 11. For all three trees, we perform the \knn{} operation after building the tree from a set of batch insertions, with a batch size of 5\% of the data set, until the entire data set is inserted. The results are shown for the 2D VisualVar data set in Figure~\ref{fig:varyk-knn-2dv} and for the 7D Uniform data set in Figure~\ref{fig:varyk-knn-7du}. In both data sets, we see that \textbf{B1} has the best \knn{} performance, followed closely by \textbf{BDL}. \textbf{B2} has significantly worse performance---this is because the construction of the tree was performed with a set of batch insertions, rather than a single construction over the entire data set, and so the tree ends up unbalanced and the \knn{} query performance suffers.

%% file: batch-kdtree/sections/experiments/plot-knn-varyk.tex
\begin{figure*}[!t]
    \begin{subfigure}{0.5\textwidth}
        \centering
        \includegraphics[width=0.8\textwidth]{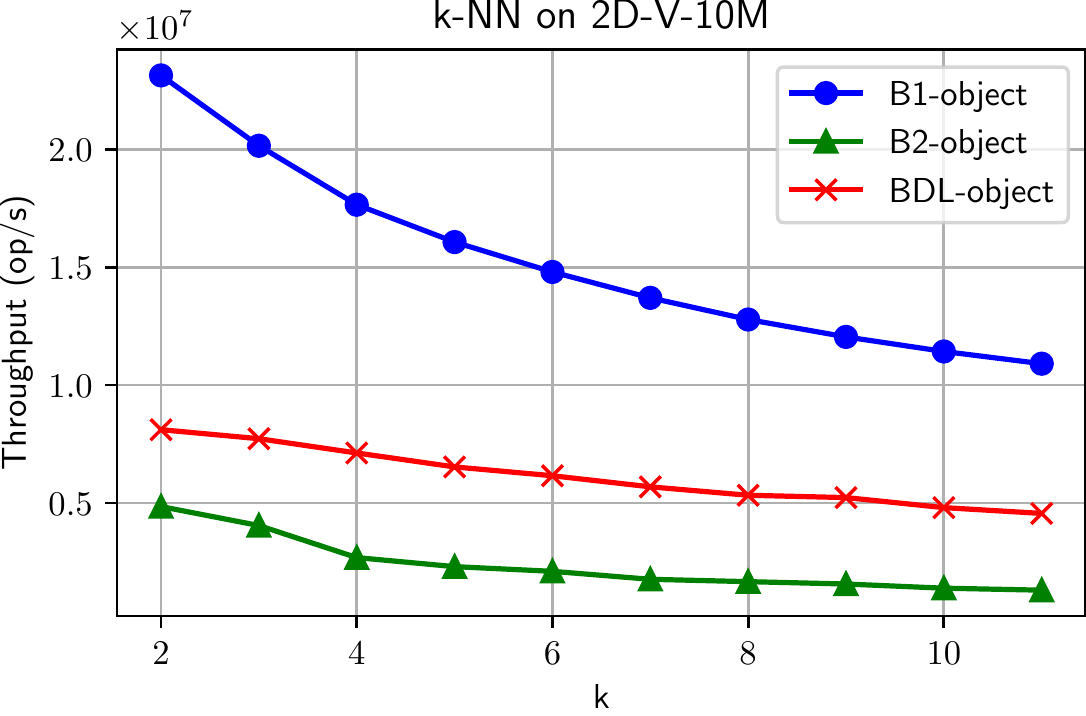}
        \caption{\knn{} on 2D-V-10M.}
        \label{fig:varyk-knn-2dv}
    \end{subfigure}
    \begin{subfigure}{0.5\textwidth}
        \centering
        \includegraphics[width=0.8\textwidth]{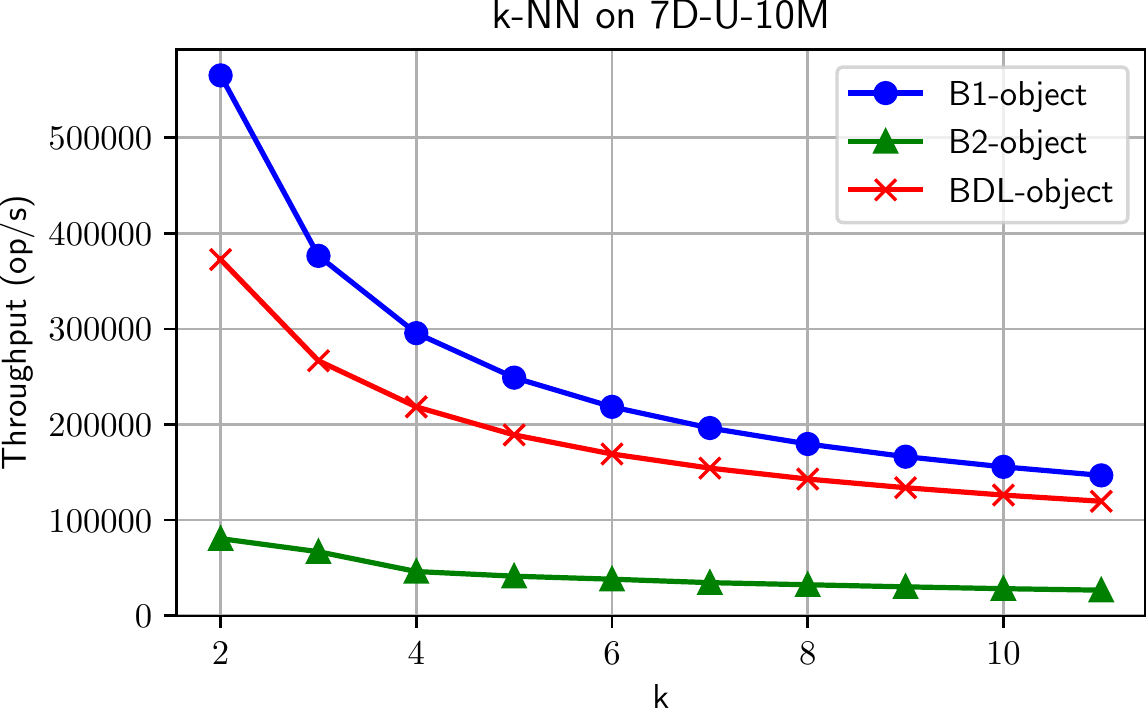}
        \caption{\knn{} on 7D-U-10M.}
        \label{fig:varyk-knn-7du}
    \end{subfigure}
    \caption{Plots of \knn{} throughput (operations per second) vs.\ $k$ using all 36 cores with hyper-threading,
    for the 2D-V-10M and 7D-U-10M data sets.}
\end{figure*}